\documentclass[reprint,superscriptaddress,aip]{revtex4-2}
\usepackage{lineno}\usepackage[normalem]{ulem}
\usepackage{xcolor}\usepackage{tikz}
\usepackage{amsmath,amssymb}
\usepackage{MnSymbol}
\usepackage{graphicx}
\usepackage{dcolumn}
\usepackage{bm}
\usepackage[sort&compress]{natbib}
\usepackage{hyperref}

\newcommand{\rT}{\rho_{\text{T}}}

\newcommand{\rH}{\rho_{\text{H}}}
\newcommand{\rSN}{\rho_{\text{SN}}}
\newcommand{\rEP}{\rho_{\text{EP}}}

\newcommand{\vP}{{\bf P}}

 \newcommand{\dx}{\text{d}{x}}

\usepackage{pstricks}

\begin{document}

\title{Instability mechanisms of repelling peak solutions in a multi-variable activator-inhibitor system}

\author{Edgar Knobloch}
 \affiliation{Department of Physics, University of California, Berkeley, California 94720, USA}

\author{Arik Yochelis}\email{yochelis@bgu.ac.il}
 \affiliation{Department of Solar Energy and Environmental Physics, Swiss Institute for Dryland Environmental and Energy Research, Blaustein Institutes for Desert Research, Ben-Gurion University of the Negev, Sede Boqer Campus, Midreshet Ben-Gurion 8499000, Israel}%
 \affiliation{Department of Physics, Ben-Gurion University of the Negev, Be'er Sheva 8410501, Israel}%

\date{\today}

\begin{abstract}
We study the linear stability properties of spatially localized single- and multi-peak states generated in a subcritical Turing bifurcation in the Meinhardt model of branching. In one spatial dimension, these states are organized in a foliated snaking structure owing to peak-peak repulsion but are shown to be all linearly unstable, with the number of unstable modes increasing with the number of peaks present. Despite this, in two spatial dimensions direct numerical simulations reveal the presence of stable single- and multi-spot states whose properties depend on the repulsion from nearby spots as well as the shape of the domain and the boundary conditions imposed thereon. Front propagation is shown to trigger the growth of new spots while destabilizing others. The results indicate that multi-variable models may support new types of behavior that are absent from typical two-variable models.
\end{abstract}

\maketitle

\noindent \textbf{Spatially localized states are a subject of intense research in many natural systems, ranging from nonlinear optics to vegetation patterns. Different formation mechanisms have been identified using simple models in one and two spatial dimensions (1D and 2D, respectively), all broadly attributed to the phenomenon of homoclinic snaking. Here, we focus on the stability properties of repulsive multi-peak states in a multi-variable activator-inhibitor model and show that their stability properties in 1D differ significantly from those in 2D. Specifically, we emphasize that while in 1D peaks are unstable they stabilize in 2D. These insights provide essential evidence that multi-variable models do not merely add to computational complexity but may support new types of behavior that is not accessible in two-variable models. This new behavior may play a significant role in studies of biological, chemical, and ecological models comprising positive feedback loops.}

\section{Introduction}

In his pioneering 1952 work~\cite{tu52} Alan Turing described a simple but general mechanism {whereby} spatially homogeneous state can become unstable to spatially periodic perturbations with an intrinsic length scale. While Turing used a two-variable parabolic partial differential equation (PDE) with fast local autocatalytic kinetics and a rapidly diffusing antagonistic agent, i.e., an activator--inhibitor (AI) system, the mechanism is general and is today also known as a \textit{finite wavenumber} instability~\cite{ch93}. The instability is broadly instrumental in pattern formation phenomena, and occurs in both nonvariational systems such as reaction--diffusion (RD) equations and variational systems, such as those described by phase field crystal (PFC) models.

When the Turing states bifurcate subcritically, i.e., in systems exhibiting bistability between a homogeneous and a pattern state, spatially localized structures embedded in a homogeneous background or holes in an otherwise periodic state, can also emerge (Ref.~\cite{knobloch2015spatial} and references therein). These localized solutions range from an isolated peak to groups of peaks organized in a `snakes-and-ladders' structure~\cite{burke2007snakes}, a consequence of so-called homoclinic snaking~\cite{woods1999heteroclinic,knobloch2015spatial}.

In a recent paper~\cite{knobloch2021stationary}, we showed that a subcritical Turing instability in Meinhardt's multi-variable AI system~\cite{meinhardt1976morphogenesis} can give rise to repelling localized states organized within a bifurcation structure called foliated homoclinic snaking. Meinhardt's model comprises four fields $A$, $H$, $S$ and $Y$ that represent, respectively, the concentrations of an activator, an inhibitor, the substrate, and a marker for differentiation~\cite{yao2007matrix}:
\begin{subequations}\label{eq:AI}
	\begin{eqnarray}
	\frac{\partial A}{\partial t}&=&c\dfrac{SA^2}{H}-\mu A+\rho_{\text{A}} Y+D_{\text{A}} \nabla^2{A}, \\
	\frac{\partial H}{\partial t}&=&cSA^2-\nu H+\rho_{\text{H}} Y+D_{\text{H}} \nabla^2{H}, \\
	\frac{\partial S}{\partial t}&=&c_0-\gamma S-\varepsilon Y S+D_{\text{S}} \nabla^2{S}, \\
	\frac{\partial Y}{\partial t}&=&d A-eY+\dfrac{Y^2}{1+fY^2}+D_{\text{Y}} \nabla^2{Y}.
    \end{eqnarray}
\end{subequations}
 {The system~\eqref{eq:AI} was suggested by Meinhardt~(\citeyear{meinhardt1976morphogenesis}) as a prototypical AI framework for studying branching. Branched filaments correspond to cells that went through irreversible differentiation (associated with the $Y$ field), such as occurs in the epithelial to mesenchymal transition. Therefore, by definition, branching requires a bistable (or binary) medium, where one of the states is in an undifferentiated state while the second state is either post-differentiation or is committed to differentiate. There are many approaches to the empirically observed branching dichotomy~\cite{metzger2008branching,hirashima2009mechanisms,menshykau2012branch,blanc2012role,celliere2012simulations,guo2014branching,guo2014mechanisms,hannezo2017unifying,xu2017turing,shan2018meshwork,zhu2018turing}, and while the explicit pattern formation mechanisms behind branching remain unclear, there is strong evidence that AI signaling dominates side-branching~\cite{sainio1997glial,tang1998ret,lebeche1999fibroblast,tang2002ureteric,gilbert2004matrix,yao2007matrix,metzger2008branching,affolter2009tissue,yao2011matrix,hagiwara2015vitro,menshykau2019image}, linking it back to Turing's morphogenesis~\cite{tu52}, i.e., to an interaction between activatory and inhibitory agents.}

 {The motivation here stems from experiments demonstrating that the growth of side-branches in the pulmonary vascular system is suppressed by excess matrix GLA protein (the $H$ field), which is an inhibitor of the bone morphogenetic protein (the $A$ field). In~\eqref{eq:AI} excess matrix GLA protein corresponds to higher values of the parameter $\rho_{\text{H}}$. Following\cite{yochelis2021nonlinear} we therefore employ $\rho_{\text{H}}$ as the key control parameter} while keeping all other parameters fixed and within the range of previous studies: $c=0.002$, $\mu=0.16$, $\rho_{\text{A}}=0.005$, $\nu=0.04$, $c_0=0.02$, $\gamma=0.02$, $\varepsilon=0.1$, $d=0.008$, $e=0.1$, $f=10$, $D_{\text{A}}=0.001$, $D_{\text{H}}=0.02$, $D_{\text{S}}=0.01$, $D_{\text{Y}}=10^{-7}$. We also note that in the original formulation of Meinhardt $D_{\text{Y}}=0$ while in~\cite{knobloch2021stationary} and likewise here $D_{\text{Y}}$ is taken as finite albeit much smaller than the smallest diffusion coefficient among all other fields, i.e., $D_{\text{Y}}\ll D_{\text{A}}$. 

\begin{figure}[tp]
\centering
    {\includegraphics[width=\columnwidth]{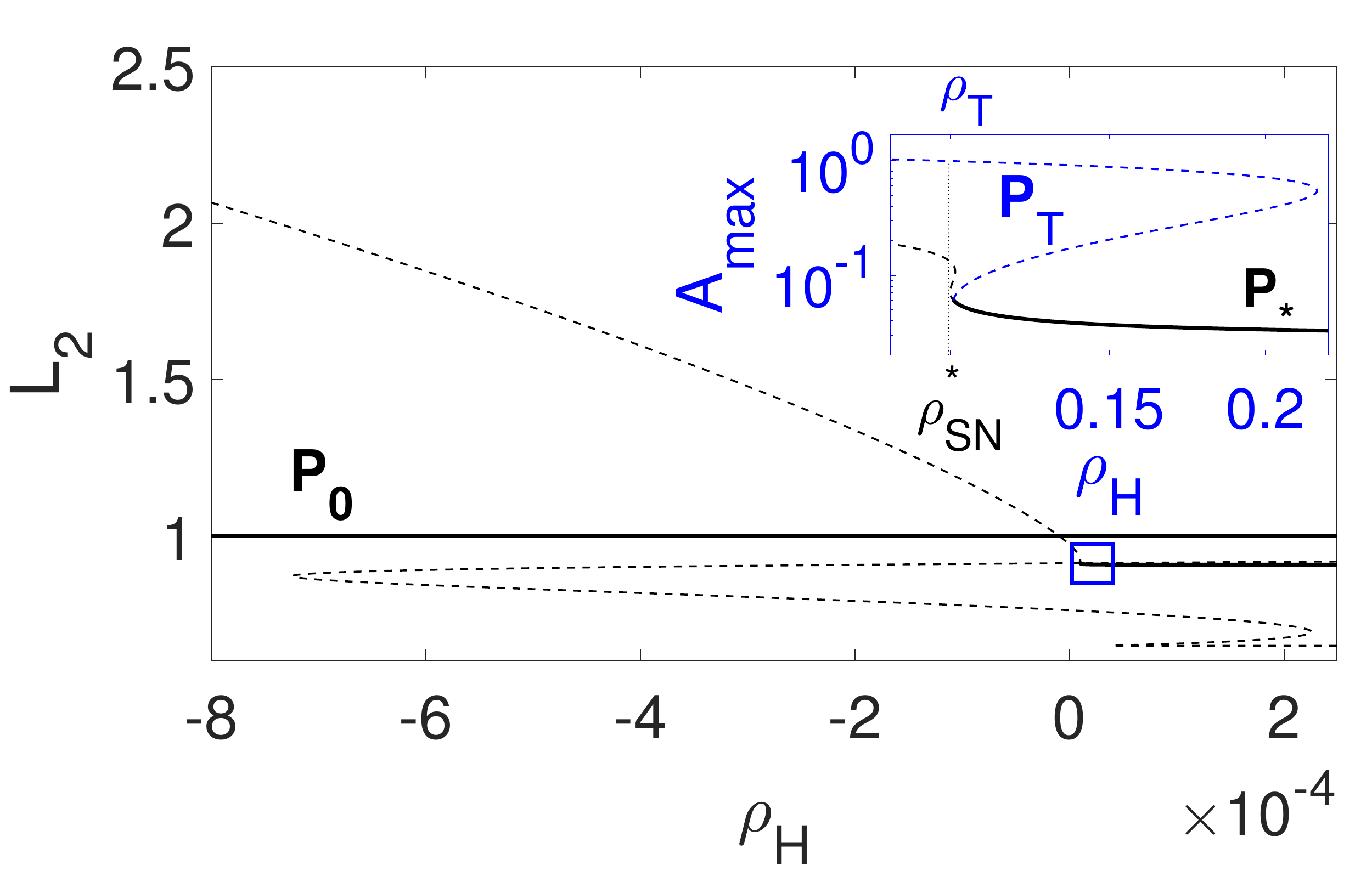}}
    \caption{Bifurcation diagram showing the coexistence of stable states ($\vP_0$ and $\vP_*$, solid black lines) together with other unstable branches in terms of the ${\text L}_2$ norm defined in~\eqref{eq:L2}, as computed from Eqs.~\eqref{eq:AI} with the parameters given in the text. The inset zooms into the rectangle in the lower left, showing the uniform and the periodic Turing states ($\vP_{\rm T}$, dashed blue line) in terms of the maximum value $A_{\max}$ of the variable $A$. The Turing instability occurs at $\rT \simeq 1.0 \cdot 10^{-5}$ while the lower left saddle node of $\vP_*$ as at $\rho_{SN}^*\simeq 0.99\cdot 10^{-5}$ (inset).}
	\label{fig:bif_uni}
\end{figure}

For the above parameter choice, system~\eqref{eq:AI} is \textit{bistable} above the Turing onset, $\rH>\rT\simeq 1.0 \cdot 10^{-5}$, with two competing spatially uniform solutions, $\vP_0\equiv (A_0,H_0,S_0,Y_0)=(0,0,c_0/\gamma,0)$ and $\vP_*\equiv (A_*,H_*,S_*,Y_*)$, both simultaneously linearly stable~\cite{yochelis2021nonlinear}. In Fig.~\ref{fig:bif_uni}, we show $\vP_0$ and $\vP_*$ (solid black lines) together with additional branches of unstable uniform solutions that are also present (dashed black lines). In addition, Fig.~\ref{fig:bif_uni} (inset) also shows the coexisting branch $\vP_{\rm T}$ of unstable spatially periodic Turing solutions with wavenumber $k_{\rm T}\simeq 2.18$ (dashed blue line) present for $\rH>\rT$. The results are shown in terms of the ${\rm L}_2$ norm
\begin{equation}\label{eq:L2}
    {\text L}_2\equiv \sqrt{L^{-1}\int_{0}^{L} \dx {\sum_{j={A,H,S,Y}} P_j^2+(\partial_x P_j)^2}}
\end{equation}
that takes into account both the solution variables and their spatial derivatives. Here $L$ is the domain length. 

\begin{figure}[ht!]
\centering
    {\fontfamily{phv}\selectfont{\large (a)}}{\includegraphics[width=0.5\textwidth]{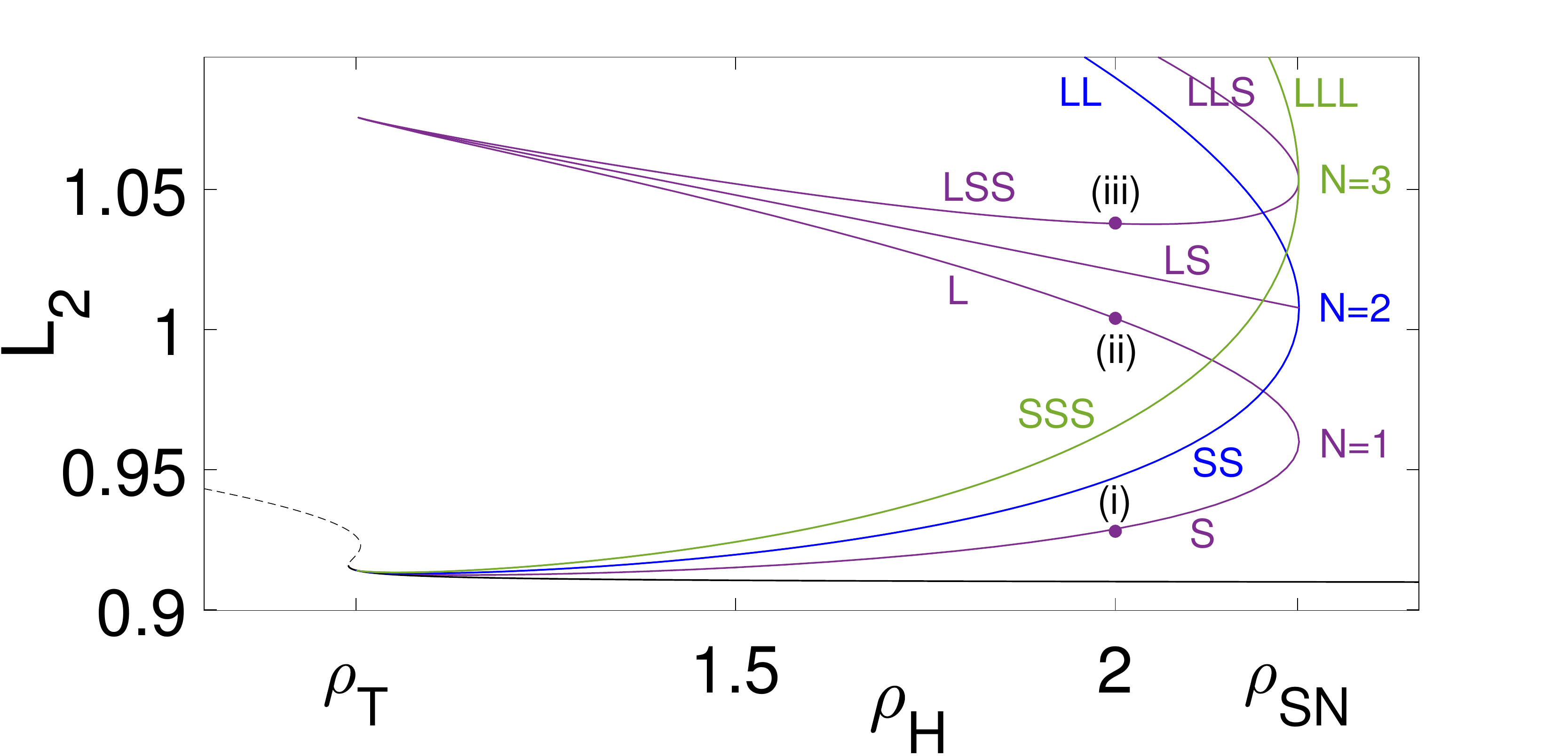}}
    {\fontfamily{phv}\selectfont{\large (b)}}{\includegraphics[width=0.4\textwidth]{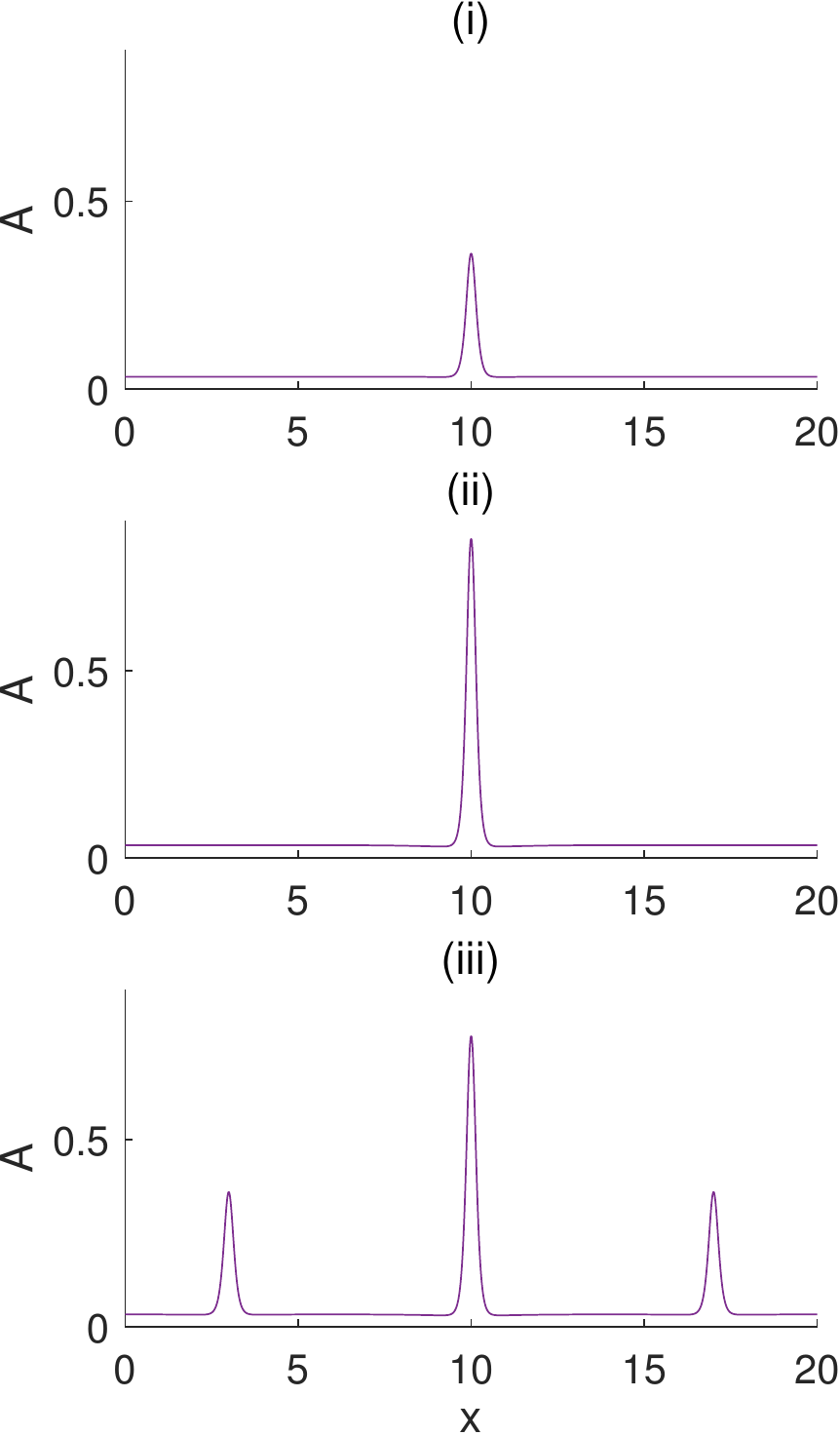}}
	\caption{(a) Bifurcation diagram showing branches of both single-peak and multi-peak solutions computed via numerical continuation on a periodic domain of length $L=20$. The labels S and L label small and large amplitude peaks, shown in panels (b-i) and (b-ii), respectively; thus LSS implies that the domain is filled with one large and two small amplitude peaks, all well separated (b-iii). The black curve at the bottom represents the spatially uniform solution $\vP_*$, with solid (dashed) lines indicating stable (unstable) spatially uniform states. In this and subsequent figures, the horizontal axis labeled $\rH$ represents $10^5\rH$. (b) Solution profiles at $\rH=2\cdot10^{-5}$ along the $N=1$ branch at locations indicated in (a). Panel (a) is reprinted from~\citet{knobloch2021stationary} with modifications to the layout.}
	\label{fig:summary}
\end{figure}

In Ref.~\cite{knobloch2021stationary}, we focused on the existence of spatially localized peak solutions in the subcritical regime $\rH>\rT$, and showed that in this regime multi-peak solutions organize in a foliated homoclinic snaking structure (Fig.~\ref{fig:summary}). In the context of system~\eqref{eq:AI}, such peaks correspond to high concentrations of activator and hence to differentiated states while the background corresponds to low activator levels and hence to undifferentiated states. These bifurcating peak solutions differ from the localized states that typically form when a uniform state becomes unstable to a Turing state, as occurs, for example, in the Gierer-Meinhardt AI model~\cite{yochelis2008front}.

Figure~\ref{fig:summary} provides a summary of the key results from Ref.~\cite{knobloch2021stationary} The figure shows that the branch of single peak solutions, $N=1$, results from a modulational or Eckhaus instability of a subcritical Turing state that appears at $\rH=\rT$. The primary $N=1$ branch extends subcritically (towards larger $\rH$) from the vicinity of the Turing instability at $\rH=\rT$. Initially the branch consists of states with one small amplitude peak (labeled S), which undergo a fold at $\rH=\rSN$ on the right and continue as a single large amplitude peak (labeled L) to the left. These L states subsequently undergo a fold close to $\rT$ in the vicinity of which the profile of the solution adds two small amplitude peaks, becoming LSS. At the same time, a branch of LS states bifurcates from the vicinity of the left fold into $\rH\gtrsim\rT$. The LSS branch subsequently extends to $\rSN$, where it connects with a primary $N=3$ branch SSS/LLL in a period-tripling bifurcation. This bifurcation is locally transcritical~\cite{knobloch2021stationary} and results in a state LLS that turns around in a nearby fold back towards smaller $\rH$. This process repeats, and the LLS connect to a fold at $\rH\simeq\rT$ where the $N=2$ state LL turns into LSSLSS (not shown); LLSS and LSLS both bifurcate nearby. These results capture the richness of the foliated snaking structure and have been obtained using numerical continuation of time-independent solutions~\cite{knobloch2021stationary}.

In Ref.~\cite{knobloch2021stationary} we conjectured that this foliated snaking structure is the result of repulsive interactions between peak solutions and that it is in fact universal. However, we also reported that time-stepping simulations appeared to indicate that all the 1D solutions are linearly unstable. In order to understand why this might be the case, we first discuss (Sec.~\ref{sec:peak_inst}) the linear stability properties of the peak solutions along the $N=1$ branch (Fig.~\ref{fig:summary}) determined by solving numerically the linear stability problem. We show that at each fold additional eigenvalues become unstable, a behavior that is rooted in the formation of additional peaks. Moreover, we show that for the majority of the large amplitude peaks, starting with the L state, the unstable eigenvalues are complex, indicating growing temporal oscillations. In Section~\ref{sec:2D} we use direct numerical simulations to show that, in contrast to the 1D case, spot solutions in 2D can be stable and study the effect of spot-spot repulsion in different 2D domains.
Finally, in the Section~\ref{sec:discussion}, we discuss the implications of this behavior for the nucleation and growth of side-branches in 2D.

\section{The instability mechanism for single-peak solutions}\label{sec:peak_inst}

In this section, we focus on the stability properties of the single-peak branch ($N=1$) and the states connected to it, i.e., the thick line in Fig.~\ref{fig:fold_evalues}. We recall that this branch is the result of a modulational or Eckhaus instability of the subcritical Turing state $\vP_{\rm T}$ that appears at $\rH=\rT$. This secondary instability occurs already at small amplitude and results in a Turing pattern whose amplitude is modulated in space with half-wavelength $L$, where $L$ is the domain size. The instability leads to two distinct states, in one of which the maximum of the modulation envelope coincides with a maximum of the Turing state while in the other it coincides with a minimum. The former evolves into an $N=1$ single peak state, while the latter evolves into a $N=2$ two-peak state, when followed in the direction of increasing $\rH$. Subsequent secondary instabilities of the Turing state lead to modulation with a shorter wavelength and these are responsible for branches with $N=3,4,\dots$ peaks. These states have been obtained via numerical continuation on periodic domains~\cite{knobloch2021stationary}, using the package AUTO.~\cite{doedel2012auto}

To calculate the stability along the $N=1$ branch, we solve the linear stability problem for different peak solutions $\vP_N(x)$, i.e.,
\begin{equation}\label{eq:eigs}
    \sigma \widetilde{\vP}(x)=\mathcal{J}{\big|}_{\vP_N(x)} \widetilde{\vP}(x),
\end{equation}
where $\sigma$ and $\widetilde{\vP}\equiv(\widetilde{A},\widetilde{H},\widetilde{S},\widetilde{Y})$ are, respectively, the temporal eigenvalue and the corresponding eigenfunction. Here $\mathcal{J}|_{\vP_N(x)}$ is the Jacobian matrix of~\eqref{eq:AI} evaluated at $\vP_N(x)$. The states $\vP_N(x)$ whose stability is of interest are all well localized in space, and have been obtained on a periodic domain of length $L=20$. For details, see Ref.~\cite{knobloch2021stationary} The stability problem is solved on the same domain and has an infinite number of discrete solutions $(\sigma,\widetilde{\vP}(x))$ for each $N$. {In the following, we compute the leading eigenvalues and associated eigenfunctions with periodic boundary conditions (PBC). Since the problem with Neumann boundary conditions (NBC) on a domain of length $L$ can be embedded in a PBC problem with period $2L$ our results also apply to the NBC problem provided perturbations with wavelength $2L$ are suppressed.}

\subsection{Single-peak states: eigenvalues along the S and L branches}

\begin{figure*}[!t]
		\centering
		{\fontfamily{phv}\selectfont{\large (a)}}\includegraphics[width=0.95\textwidth]{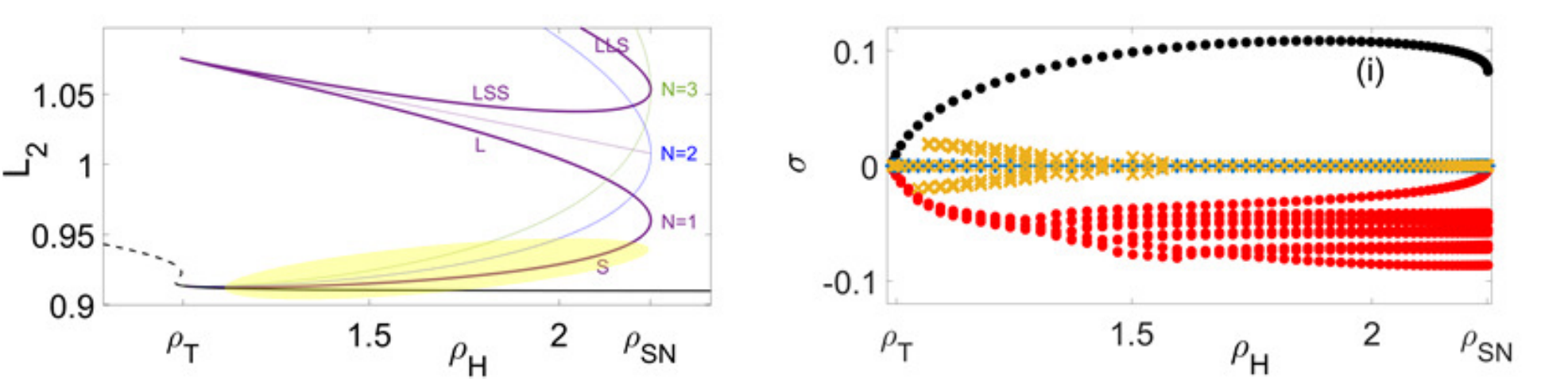}
		{\fontfamily{phv}\selectfont{\large (b)}}\includegraphics[width=0.95\textwidth]{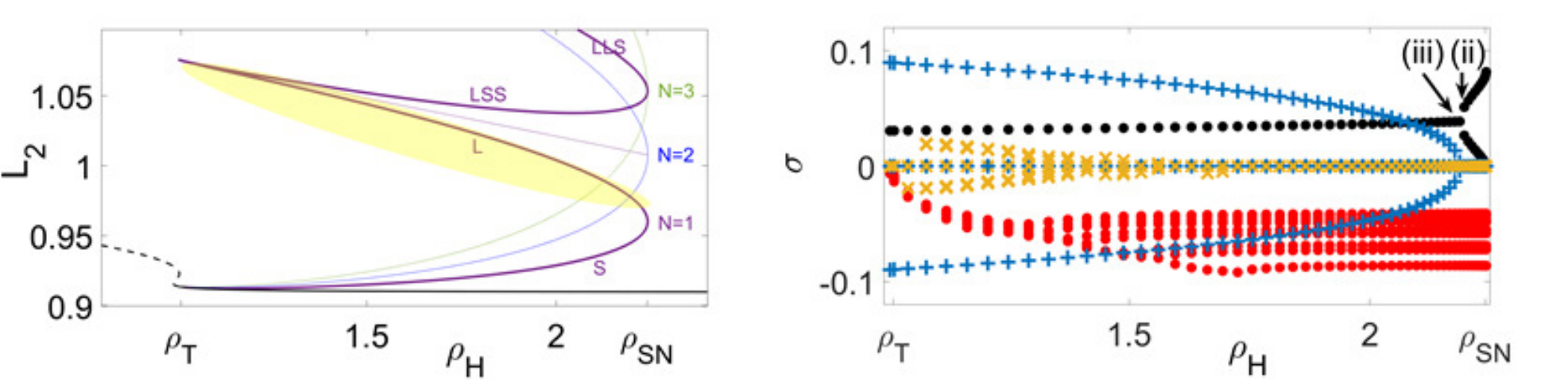}
		{\fontfamily{phv}\selectfont{\large (c)}}\includegraphics[width=0.95\textwidth]{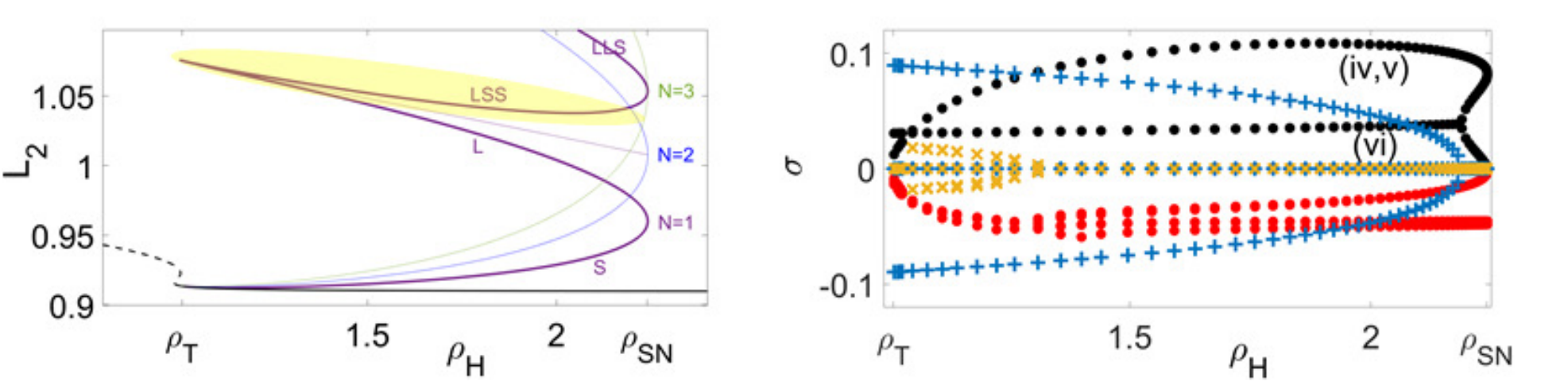}
		{\fontfamily{phv}\selectfont{\large (d)}}\includegraphics[width=0.95\textwidth]{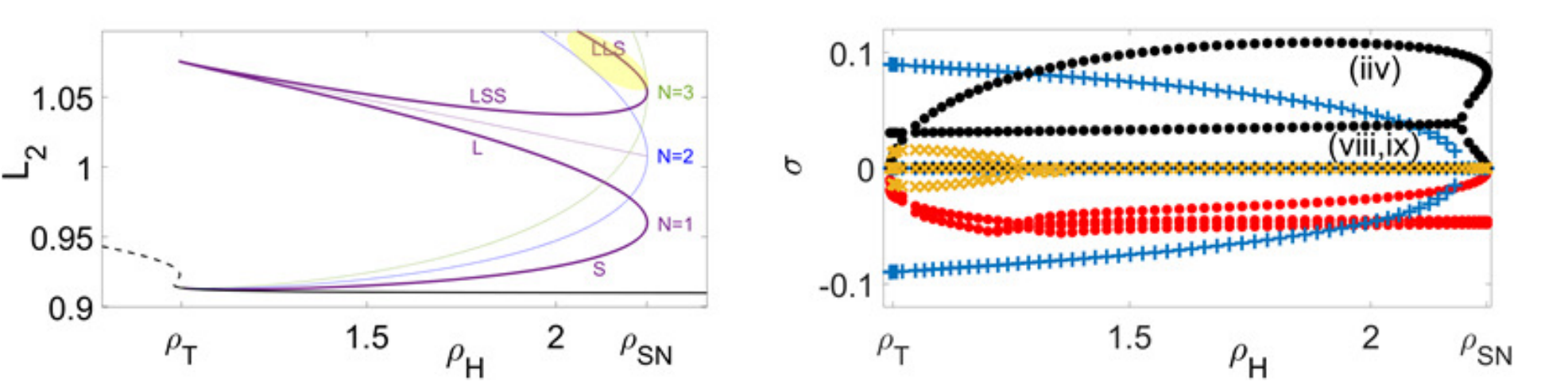}
		\caption{Solutions along the primary $N=1$ branch (left panels, yellow highlight): (a) S branch, (b) L branch, (c) LSS branch, and (d) LLS branch. Corresponding temporal eigenvalues $\sigma$ are shown in the right panels. Twelve eigenvalues with the largest real parts (marked by $\bullet$) are computed, with black (red) color indicating Re$[\sigma]>0$ (Re$[\sigma]<0$); the corresponding imaginary parts are shown using blue $+$ (orange $\times$) symbols, respectively.}
		\label{fig:fold_evalues}
\end{figure*}

\begin{figure*}[!t]
		\centering\includegraphics[width=1\textwidth]{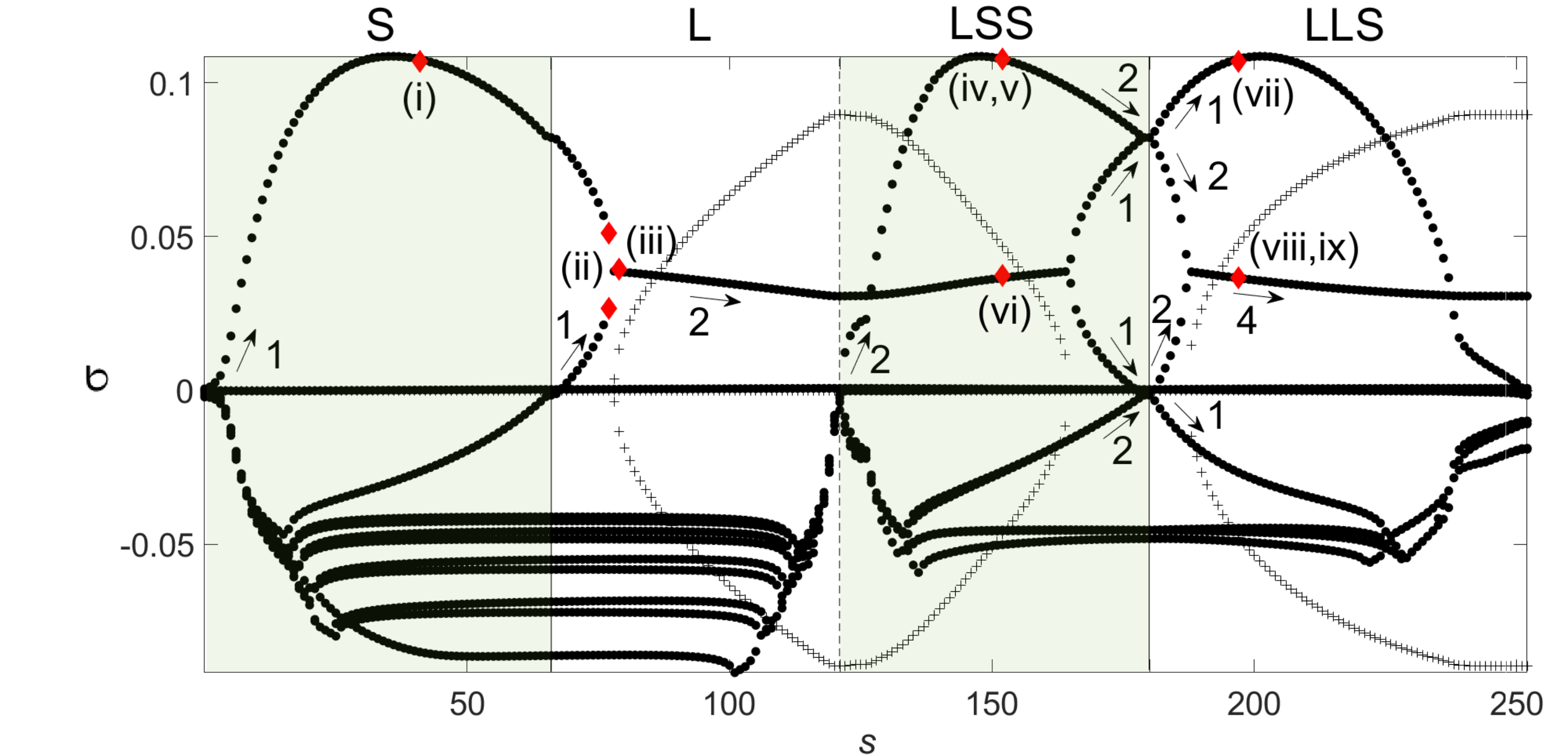}
		\caption{Temporal eigenvalues $\sigma$ as a function of the arclength $s$ along the $N=1$ branch showing the S, L, LSS, and LLS branches bach-to-back. Integers indicate the eigenvalue multiplicity; multiple eigenvalues indicate degeneracy arising from widely separated, exponentially localized peaks along the LSS and LLS branches. The labeled red diamonds indicate the locations corresponding to the eigenfunctions shown in Fig.~\ref{fig:fold_evalues_prof}. See Fig.~\ref{fig:fold_evalues} for additional details.}
		\label{fig:fold_evalues_cont}
\end{figure*}

\begin{figure*}[ht!]
		\centering
		{\fontfamily{phv}\selectfont{\large (a)}}\includegraphics[width=0.95\textwidth]{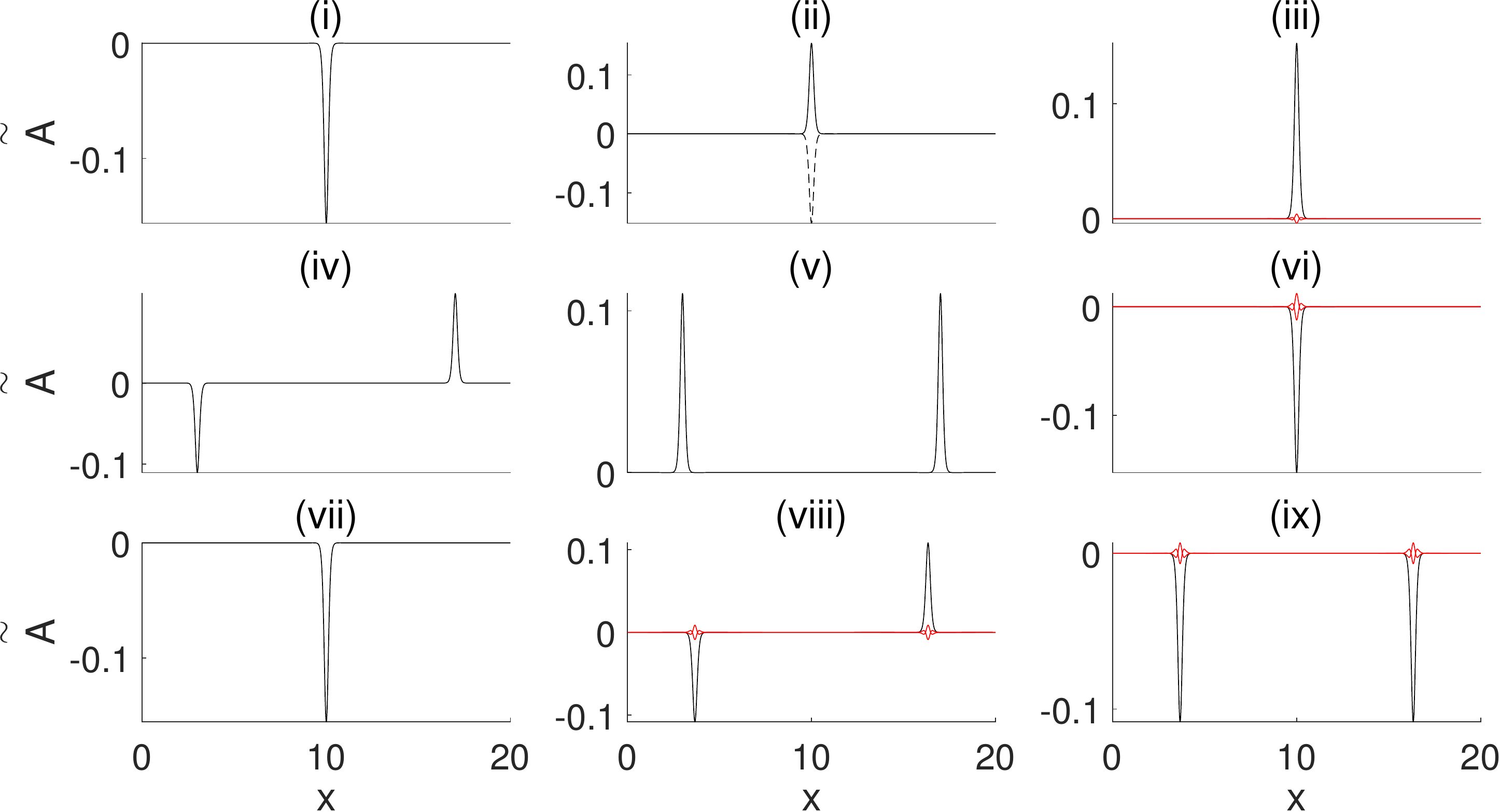}\vskip 0.2in
		{\fontfamily{phv}\selectfont{\large (b)}}\includegraphics[width=0.95\textwidth]{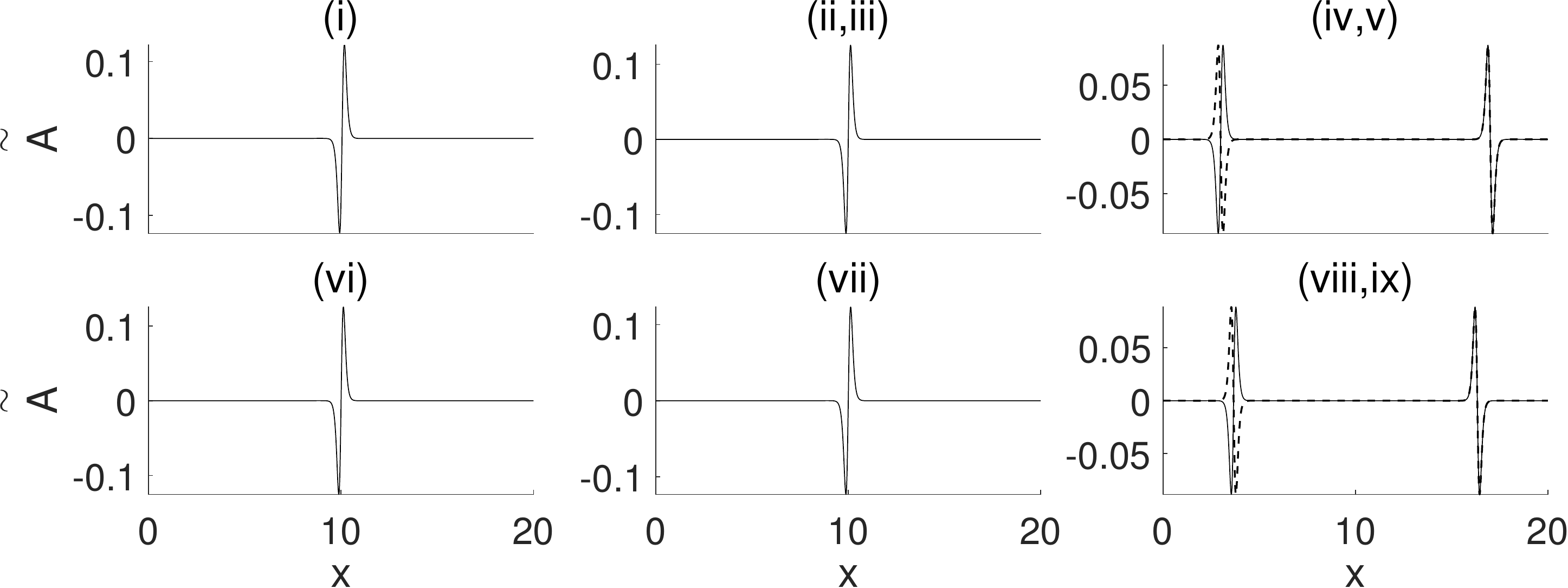}
		\caption{(a) The $\widetilde{A}$ component of the normalized eigenfunction $\widetilde \vP(x)$ computed on a periodic domain with period $L=20$ by solving the eigenvalue problem~\eqref{eq:eigs} for the eigenvalues at locations indicated in the right panels (a-d) in Fig.~\ref{fig:fold_evalues}, where (i) and (iv)-(ix) are at $\rH\simeq 2.0 \cdot 10^{-5}$ while (ii) and (iii) are at $\rH\simeq 2.2 \cdot 10^{-5}$ and $\rH\simeq 2.18 \cdot 10^{-5}$, respectively; see also Fig.~\ref{fig:fold_evalues_cont}. The dashed (solid) lines in (ii) indicate eigenfunctions that are associated with the largest (next largest) eigenvalues. The red lines in (iii), (vi), (viii), (ix) correspond to the imaginary parts of complex eigenfunctions whose real part is indicated in black. (b) Neutral eigenfunctions corresponding to translation modes in (a). Each eigenfunction is defined to within a sign. The computations were done using the MATLAB function \textit{eigs} with Euclidean norm.}
		\label{fig:fold_evalues_prof}
\end{figure*}

We start at small amplitude solutions and compute numerically the first 12 temporal eigenvalues $\sigma$ of $\vP_1(x)$ with the largest real parts. This calculation indicates that the S states are once unstable (right panel in Fig.~\ref{fig:fold_evalues}(a)), a fact consistent with the subcritical branching direction. Note that this single unstable eigenvalue is real (Fig.~\ref{fig:fold_evalues_cont})) and represents an amplitude mode, as revealed by the eigenfunctions $\widetilde{A}$ shown in Fig.~\ref{fig:fold_evalues_prof}(a-i) at $\rH \simeq 2.0 \cdot 10^{-5}$. In addition, a typical neutral translation mode is also present (Fig.~\ref{fig:fold_evalues_prof}(b-i)).

As one approaches the right fold on the $N=1$ branch where the S state turns into an L state (at $\rSN\simeq 2.242 \cdot 10^{-5}$), a real negative eigenvalue approaches zero and passes through it at the fold (right panel in Fig.~\ref{fig:fold_evalues}(a)). As a result above the right fold, the $N=1$ states now possess two real positive eigenvalues (right panel in Fig.~\ref{fig:fold_evalues}(b)); these approach one another as the L branch is followed to the left and collide at $\rH\simeq 2.19 \cdot 10^{-5}$, becoming complex for smaller $\rH$. This situation extends all the way to $\rT$. Figure \ref{fig:fold_evalues_prof}(a-ii) shows the real eigenfunctions at $\rH\simeq 2.2 \cdot 10^{-5}$ while panel (a-iii) shows the complex eigenfunction at $\rH\simeq \rH=2.18 \cdot 10^{-5}$, i.e., just after the coalescence of the two real eigenvalues; a neutral translation mode is also present, just as for S, but is not shown. Note that near the right fold the L state has two distinct real positive eigenvalues, one originating in the subcritical instability and so inherited from the instability of S and the other, smaller, eigenvalue associated with the proximity to the fold. It is the latter that is responsible for the transformation from S to L. The corresponding eigenfunctions $\widetilde{A}$ are shown in panel (a-ii), the dashed (solid) curve corresponding to the larger (smaller) positive eigenvalue, and are of amplitude type.

\subsection{Multi-peak states: eigenvalues along the LSS and LLS branches}

At the left fold of the L branch, two additional real eigenvalues pass through zero (right panel in Fig.~\ref{fig:fold_evalues}(c)), becoming unstable. These eigenvalues increase beyond the fold, crossing the real part of the complex eigenvalues inherited from the L branch, and become the most unstable modes along LSS (Fig.~\ref{fig:fold_evalues}(c)). The transition from a dominant oscillatory instability to a monotonically growing instability as one proceeds from the left fold to the next right fold appears to coincide with a similar transition among the spatial eigenvalues of $\vP_*$ that occurs at the \textit{exchange point} $\rH=\rEP\simeq 1.04 \cdot 10^{-5}$, cf. Ref.~\cite{knobloch2021stationary}. The reason for this remains unclear.

The presence of two new and nearly identical unstable eigenvalues above the left fold is associated with the bifurcation to the LS branch that takes place very close to the left fold. The complex eigenvalues subsequently collide on the real axis (at $\rH\simeq 2.19\cdot 10^{-5}$, see Fig.~\ref{fig:fold_evalues}(c)) and split into two real positive eigenvalues, one of which decreases to zero at the right fold while the other continues to increase. Thus near the right fold the LSS branch has {\it four} real positive eigenvalues (cf. Fig.~\ref{fig:fold_evalues_cont}). Figure~\ref{fig:fold_evalues_prof}(a) shows the two real and one complex eigenfunction at $\rH\simeq 2\cdot 10^{-5}$ (panels (a-iv)--(a-vi)), i.e., before the collision of the complex eigenvalues. As expected, the profiles show that the complex eigenvalue is associated with the L peak, just as it is along the L branch. In contrast, the real eigenfunctions are associated with the real positive eigenvalues generated at the left fold, and these are in turn responsible for the creation of the two smaller peaks SS on the LSS branch. We see that these eigenfunctions correspond to odd and even modes, eigenfunction (a-iv) tending to increase the amplitude of one of the S peaks while suppressing the amplitude of the other S peak, while (a-v) enhances (or suppresses) both.  Moreover, since the SS peaks are identical and exponentially localized whenever $\rH>\rEP$, the two eigenfunctions can be combined into two new eigenfunctions, one of which is localized at the first S peak while the other is localized at the second S peak. Thus the stability problem generates three distinct eigenfunctions each of which is associated with one of the peaks in the LSS solution. As one approaches the right fold all three peaks approach the same height and become equispaced, and the LSS/LLS branch connects to the $N=3$ branch just above its own fold and very close to the fold on the LSS/LLS branch (Fig.~\ref{fig:summary}). As already mentioned, this bifurcation is locally transcritical~\cite{knobloch2021stationary}. 

Along LSS there are, in addition, three vanishing eigenvalues; the corresponding eigenfunctions are shown in Fig.~\ref{fig:fold_evalues_prof}(b), panels (iv)--(vi). These are readily seen to be associated with, effectively independent, translations of the three peaks LSS. This is because the translation modes are given by $\text{d}( \Re\text{e}\, A)/\text{d}x$. We refer to the former two as ``double'' modes and to the latter as a ``single'' mode. In fact, there is only one exact translation mode, with an eigenvalue that is exactly zero, corresponding to the simultaneous translation of all three peaks by the same amount. The two ``double'' modes are phase modes associated with the SS peaks, with (b-v) associated with the translation of SS in the same direction while L remains at rest, and (b-iv) associated with the translation of SS in opposite directions while L again remains at rest (Fig.~\ref{fig:fold_evalues_prof}(b)). Once again, we can add and subtract (b-iv) and (b-v) obtaining, to exponential accuracy, pure translation modes for one or other S peak. Thus there are in effect three independent translation modes, a consequence again of the exponential localization of each peak in the solution (cf. Fig.~\ref{fig:fold_evalues_prof}(b)).

At the right fold, two new positive eigenvalues appear, at the same time as the small LSS eigenvalue passes through zero and becomes stable ($\rH\simeq 2.19 \cdot 10^{-5}$, see Fig.~\ref{fig:fold_evalues}(d)). The former are almost identical because they are associated with symmetric and antisymmetric eigenfunctions localized at the two S peaks that are well separated, while the latter is associated with an eigenfunction that is localized at L. Of the former the antisymmetric eigenfunction is associated with the LSS/LLS fold while the symmetric eigenfunction is associated with the connection to the $N=3$ branch. At the same time, the intermediate real positive LSS eigenvalue crosses the pair of large positive eigenvalues (the kink in the eigenvalue curves associated with this crossing (Fig.~\ref{fig:fold_evalues_cont}) is a consequence of resonance), resulting in one large and positive eigenvalue along LLS together with two pairs of smaller real positive eigenvalues. Thus the presence of the fold between LSS and LLS is associated with a net increase of one unstable mode. Following the two pairs of almost identical eigenvalues, we see that they too collide on the real positive axis, and subsequently split, forming two pairs of almost identical complex eigenvalues. The real part of these eigenvalues remains almost constant, much as was the case for the first complex eigenvalue along the L branch.

Following the LLS branch further, we see that the single large positive eigenvalue starts to decrease, eventually crossing the real part of the complex eigenvalues, thereby heralding a transition from a dominant monotonic instability to a dominant oscillatory instability. The real eigenvalue subsequently stabilizes at the LLS left fold, simultaneously with the appearance of additional unstable eigenvalues, just as before. As shown in Fig.~\ref{fig:fold_evalues_prof}(b), panels (vii)-(ix), the three approximately independent translation modes persist along LLS, as expected from the fact that the LLS branch also consists of three exponentially localized peaks.

\subsection{Subsequent behavior}

Observe that the behavior along the LSS branch reported above resembles that along the S branch. Except for multiplicity the (large) real eigenvalues behave in almost the same way. In each case they collide with new unstable eigenvalues that arise at the respective right folds and form a pair of complex eigenvalues, again up to multiplicity. The real parts of these complex eigenvalues are likewise similar and almost independent of the location. As a result the main difference between the behavior along the L and LLS branches is the presence of the large eigenvalue inherited from the splitting of the complex eigenvalues along L/LSS. We conjecture that this behavior repeats in the same fashion as one follows the $N=1$ branch higher and higher along the foliated snaking structure, and moreover that similar behavior takes place along the $N>1$ branches in Fig.~\ref{fig:summary}.

This observation is significant because there are currently two generic bifurcation structures that are known to organize spatially localized structures on the line. In standard snaking, exemplified by the Swift-Hohenberg equation with quadratic-cubic nonlinearity, the localized solutions grow by adding new wavelengths symmetrically on either side. Thus growth of the structure is associated with instability of the fronts at either end, and such growth is independent of the length of the structure. As a result when one computes the linear stability of the different solutions one finds that it is always determined by the same two leading eigenvalues, one corresponding to a symmetric mode and the other to an antisymmetric mode. As shown in Ref.~\cite{burke2006localized}, these eigenvalues oscillate around zero, asymptotically in phase, passing together through zero at every fold. Thus the stability of a localized solution switches periodically, from unstable to stable to unstable etc., as one passes up the snaking structure. Multiple coexisting stable structures are the result.

There is, however, a second possibility, first identified in Ref.~\cite{beaume2018three}. In this mechanism, the presence of a fold is always associated with a {\it new} instability mode. Thus as one proceeds up the snaking structure, the solutions acquire more and more unstable modes, and these new modes of instability arise as a consequence of the growth in length or equivalently the number of units within the structure.

In the present case, the localized structures are organized in a foliated snaking structure, but the stability of the solutions resembles the situation identified in Ref.~\cite{beaume2018three}. In other words, as a solution acquires new degrees of freedom via the nucleation of new peaks in the domain, it also acquires every time a new mode of instability, leading to increasingly unstable states as the number of peaks in the solution increases.

We have checked the above results using direct numerical integration (DNS). These confirm that both single-peak and multi-peak solutions are indeed unstable to small perturbations, and that the instability always results in decay to the stable $\vP_*$ state. This absence of stability of the peak states is a consequence of amplitude instability. In the next section, we show that in two spatial dimensions (2D) the stability situation is quite different.

\section{Stability in two spatial dimensions}\label{sec:2D}

In two dimensions we do not have the corresponding bifurcation diagrams and moreover, a linear stability analysis of the type performed in the previous section becomes much more involved. Consequently, we turn to DNS of the system (\ref{eq:AI}). We solve the equations in a rectangular domain $0\le x\le 20$, $0\le y\le 40$ with periodic boundary conditions (PBC) or Neumann boundary conditions (NBC) and employ the same parameter values as used earlier. These computations demonstrate that in 2D single-peak states can be stable and confirm the conjectured peak-peak repulsion.
\begin{figure}[tp!]
\centering
    {\fontfamily{phv}\selectfont{\large (a)}}{\includegraphics[width=\columnwidth]{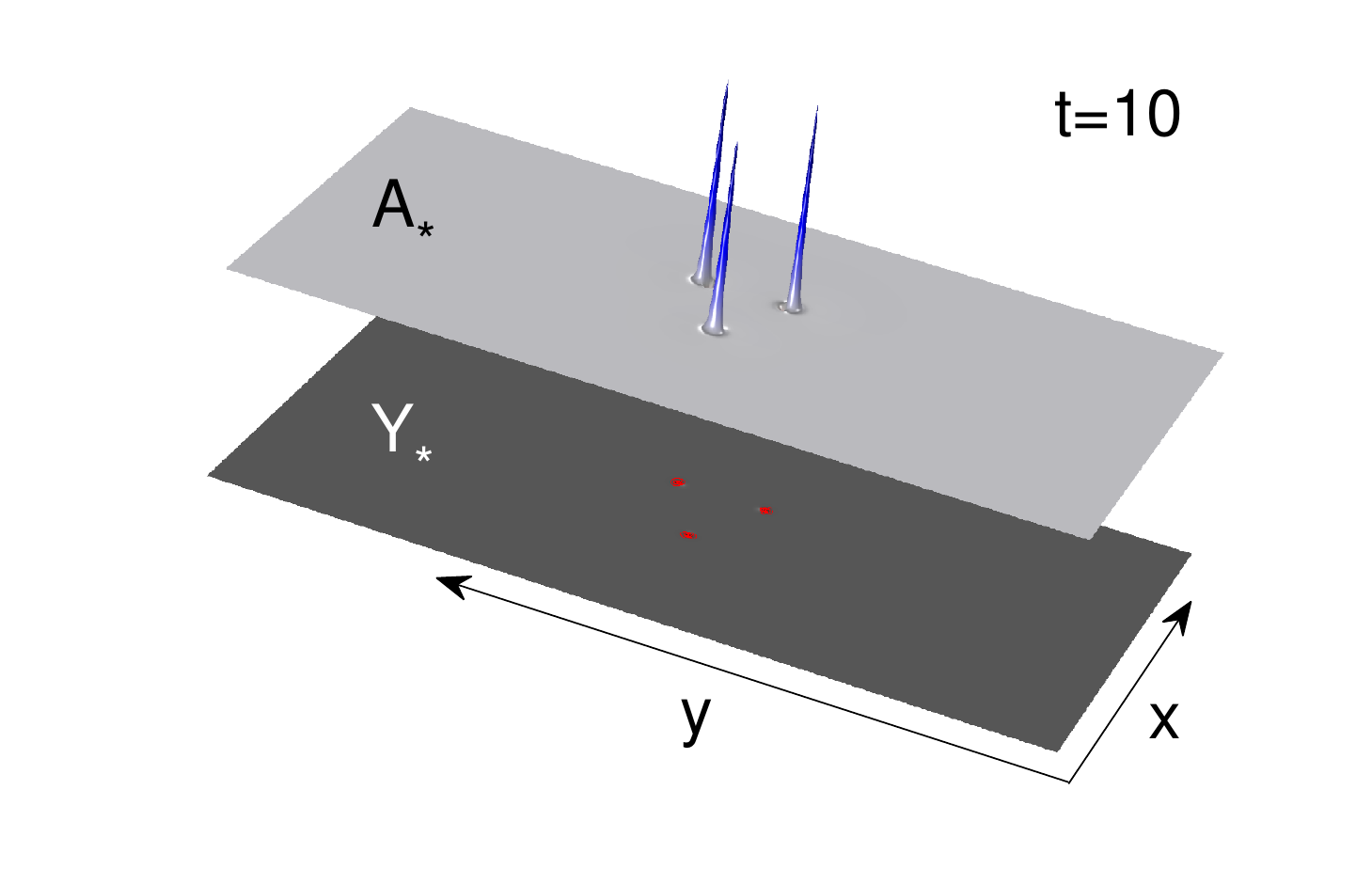}}
    {\fontfamily{phv}\selectfont{\large (b)}}{\includegraphics[width=\columnwidth]{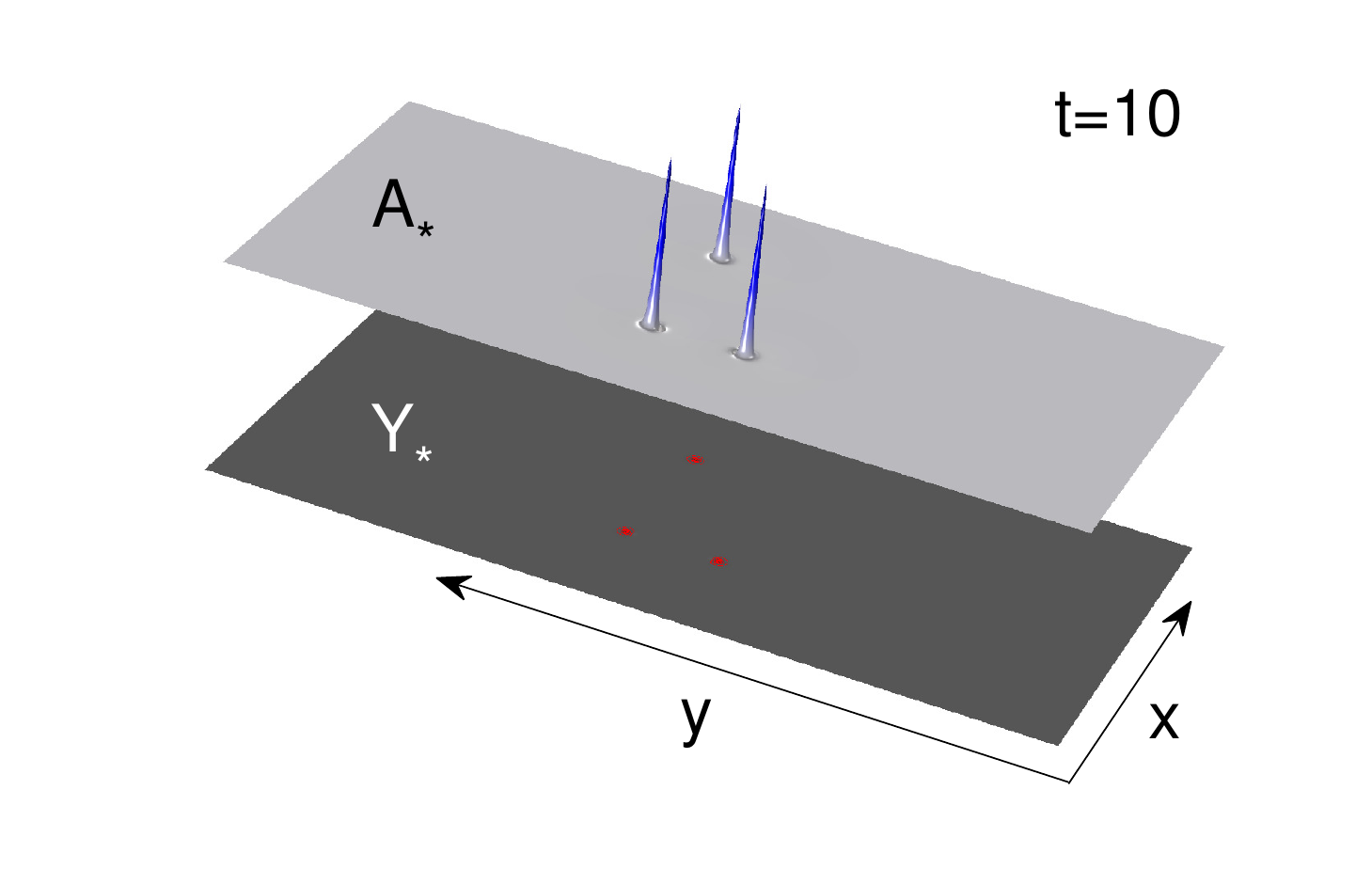}}
    {\fontfamily{phv}\selectfont{\large (c)}}{\includegraphics[width=\columnwidth]{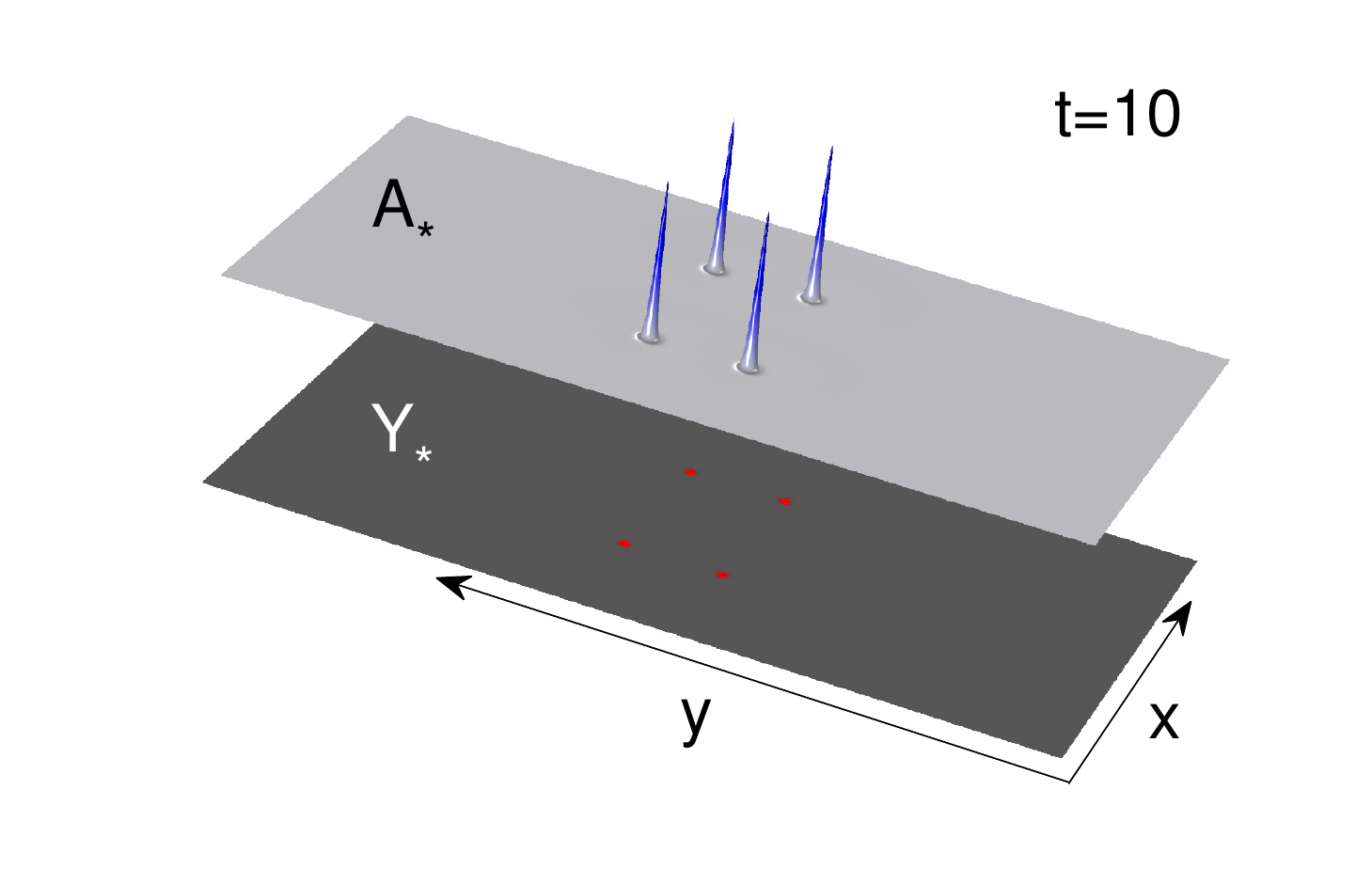}}
    \caption{Three different configurations of stable spots or pulses at early times, obtained from a direct numerical simulation of~\eqref{eq:AI} at $\rH=2.0\cdot 10^{-5}$ with periodic boundary conditions in $x\in[0,20]$ and Neumann boundary conditions in $y\in[0,40]$. The $A$ field is shown on top with the $Y$ field shown below and the red contours marking the peak locations; dark colors indicate higher values of the field. Figure~\ref{fig:locations} shows the result of evolving these states over long timescales and with different combinations of boundary conditions.}
\label{fig:2D_3peaks}
\end{figure}

\begin{figure*}[tp]
\centering
    {\fontfamily{phv}\selectfont{\large (a)}}{\includegraphics[width=0.9\textwidth]{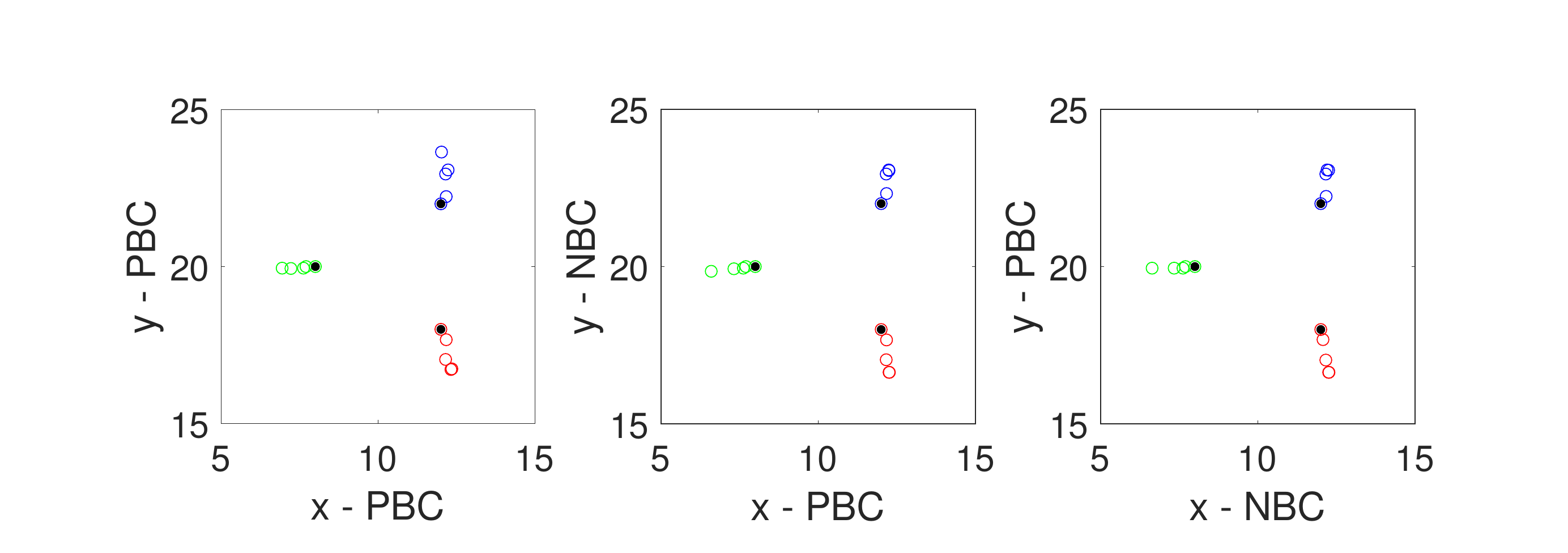}}
    {\fontfamily{phv}\selectfont{\large (b)}}{\includegraphics[width=0.9\textwidth]{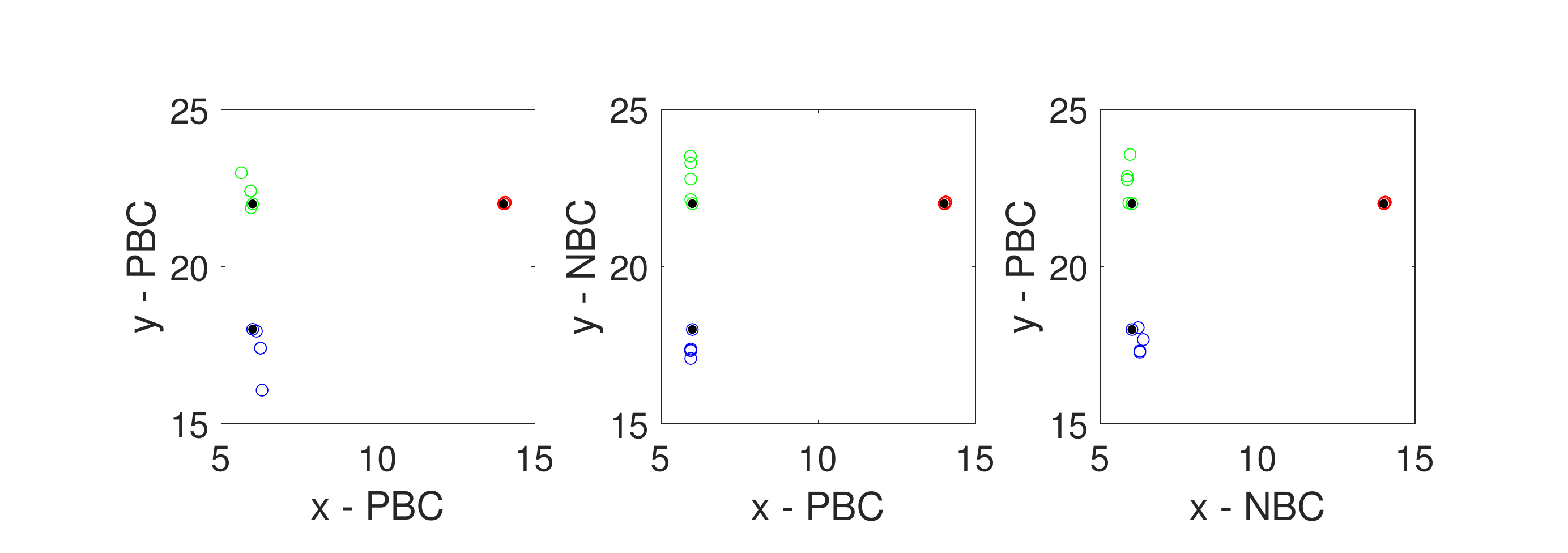}}
    {\fontfamily{phv}\selectfont{\large (c)}}{\includegraphics[width=0.9\textwidth]{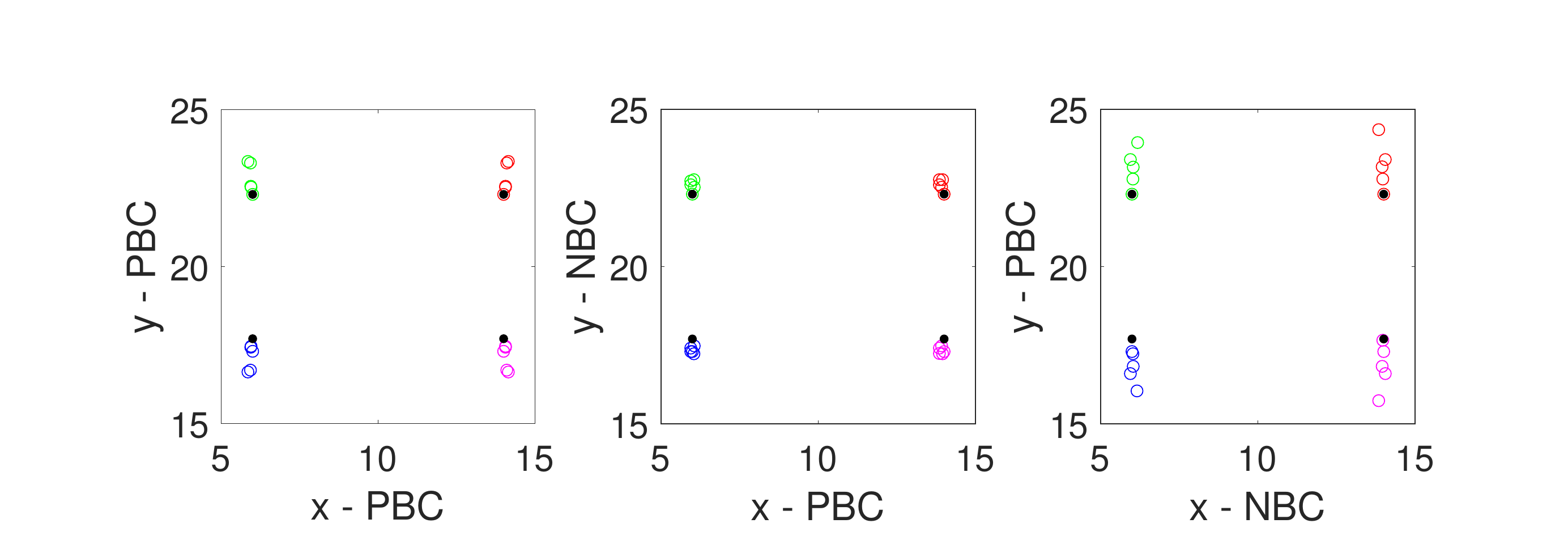}}
    \caption{Repulsive dynamics of three different configurations of stable spots in the $(x,y)$ plane, obtained from direct numerical simulation of~\eqref{eq:AI} at $\rH=2.0\cdot 10^{-5}$ with various combinations of periodic boundary conditions (PBC) and Neumann boundary conditions (NBC) along the domain boundary as indicated in the panels, starting from the three configurations shown in Fig.~\ref{fig:2D_3peaks}, indicated here by the black dots. The snapshots are taken at times $t=2\cdot10^5$, $5\cdot10^5$, $1\cdot10^6$, and $3\cdot10^6$, with other parameters as in Fig.~\ref{fig:2D_3peaks}. Only a portion of the $(x,y)\in [0,20]\times[0,40]$ domain is shown in each case.}
\label{fig:locations}
\end{figure*}

\begin{figure*}[tp]
\centering
    {\includegraphics[width=0.9\textwidth]{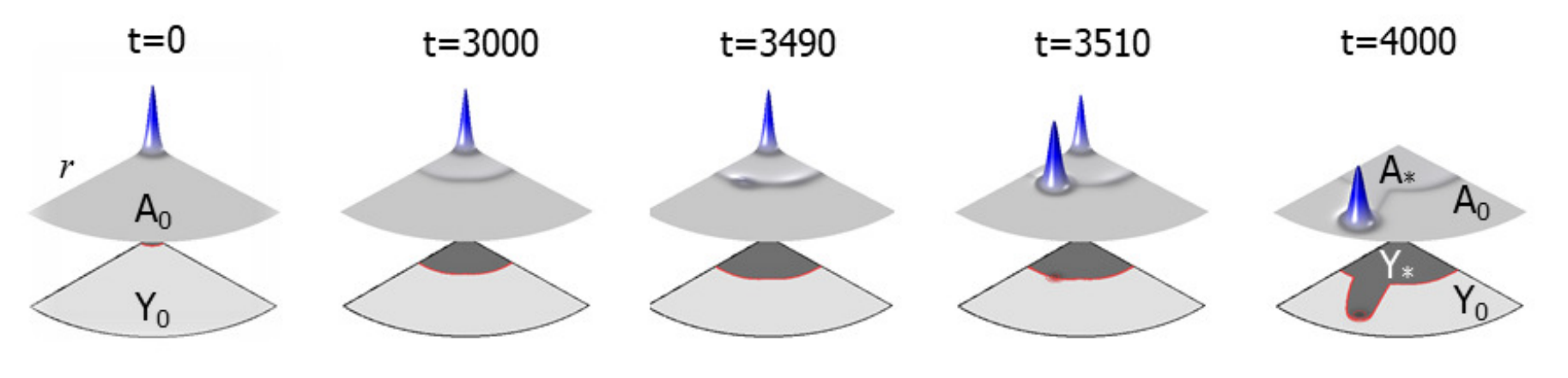}}
    \caption{Snapshots showing the evolution of an initial spot embedded in the background of $A_0$ as obtained from a direct numerical simulation of~\eqref{eq:AI} at $\rH=2.0\cdot 10^{-5}$, on a quarter of a circular domain of radius $R=4$ and Neumann boundary conditions everywhere on the boundary. The $A$ field is shown on top with the $Y$ field shown below and the red contours marking the peak locations; dark colors indicate higher values of the field.}
\label{fig:peak_front}
\end{figure*}

In 2D we may simply extend the existing 1D states in the $y$ direction, generating stripe-like localized structures. Such structures inherit the instability properties from the 1D analysis, although they may also be unstable to modes with a nonzero wavenumber $k_y$ in the $y$-direction resulting in transverse (zigzag) and varicose (breakup) instabilities~\cite{kolokolnikov2006stability,kolokolnikov2006zigzag,yochelis2008formation}. Thus, such 2D stripes are at least as unstable as in 1D and it is, therefore, no surprise that these states also decay into the $\vP_*$ state. However, in 2D we can also find peak states in the form of a two-dimensional axisymmetric structure that may properly be called a spot. Such spot-like states are common in AI systems in the plane and have been widely studied as an example of particle-like interactions~\cite{pearson1993complex,purwins2005dissipative,vanag2007localized,chen2011stability,jamieson2016localized,wong2021spot}.

Our computations show that in 2D spots may be stable and that they can exist even above the foliated snaking region, i.e., for $\rT<\rH<5.0 \cdot 10^{-5}$. Like the corresponding 1D structures, the spots are embedded in a $\vP_*$ background and exhibit a profile $\vP(r)$ that decays monotonically to $\vP_*$ (i.e., without any spatial oscillations), as shown in Fig.~\ref{fig:2D_3peaks}. Moreover, such spots exhibit repulsive interaction, once two or more of them are placed in adjacent locations, as shown for different sets of three and four spots in Fig.~\ref{fig:locations}. We emphasize that Fig.~\ref{fig:locations} represents snapshots of the spot configurations that continue to evolve albeit ever more slowly as the inter-spot separation increases.

Figure~\ref{fig:2D_3peaks} shows the early time evolution of three different initial conditions consisting of three and four identical stable peaks, forming (a) an isosceles triangle with the base parallel to the $x=20$ boundary, (b) a right-angle triangle with one side parallel to the $x=0$ boundary, and (c) a rectangle with sides parallel to the boundaries. In Fig.~\ref{fig:locations}, we show the corresponding longtime evolution of these states obtained via DNS with various combinations of PBC and NBC along the domain boundary. In general, with PBC in both $x$ and $y$ directions the boundaries do not exert any additional forces on the configuration and the spots repel one another with a repulsive force that depends on their separation, as shown best in Fig.~\ref{fig:locations}(a). In contrast, if the boundary conditions at $x=0$ and $x=20$ are replaced by NBC ($x$-NBC) an isolated spot will interact with image spots, leading to an additional force on the spot repelling it from the boundary (as best shown in Fig.~\ref{fig:locations}(b,c)); under the action of this force an isolated spot would move to the center of the rectangular domain at $x=10$ where this force vanishes. Thus, the imposed boundary conditions play a key role. A similar force is present if the boundary conditions at $y=0$ and $y=40$ are replaced by NBC ($y$-NBC), but in this case, these forces are much weaker.

While the effect of the boundary conditions is not so prominent in the case of the isosceles configuration (Fig.~\ref{fig:locations}(a)) it becomes evident for an initial condition in the form of a right-angled triangle (Fig.~\ref{fig:locations}(b)). In this case, the forces are dominated by repulsion between the pair of spots parallel to $x=0$ regardless of the boundary conditions since these peaks are closer to one another than to any image peak; with NBC along $y=0,40$ (y-NBC), we expect that the peaks will come to rest at $y=10,30$, where the $y$-force vanishes, but the force driving the system to this state is so weak that this state is effectively unreachable. Moreover, before this happens the $y$-force is superseded by a stronger $x$-force, due to the third spot (red), although this force is so week that this spot barely moves. Likewise, with NBC at $x=0,20$ (x-NBC), the pair of spots nearest to $x=0$ first separate in the $y$ direction, but eventually the $x$-force from both the red spot and from the interaction of these spots with their image spots under reflection, primarily in $x=0$, begins to dominate. Both $x-$ and $y$-forces progressively deform the triangle, resulting over time in a slight reduction in the right angle. The qualitative impact of the boundary conditions persists for the rectangular arrangement of four spots (Fig.~\ref{fig:locations}(c)). In the case of PBC, the spots parallel to the $x=0,20$ boundaries drift apart, mostly in the $y$ direction, since their separation in this direction is least. In contrast, with NBC at $y=0,40$ (y-NBC) this expansion is reduced by the presence of image spots in the $y$-direction (second panel). These image spots are absent when x-NBC are imposed instead, allowing for faster expansion in this direction (third panel).

The stability of isolated 2D spots required for the above study only holds when the spots are embedded in the uniform background $A_*$. However, as has been already mentioned here and in Ref.~\cite{yochelis2021nonlinear}, in this range of parameters front solutions connecting $A_*$ and $A_0$ can also exist, see the bistability region in Fig.~\ref{fig:bif_uni}. These fronts always propagate such that $A_*$ invades $A_0$ (not shown); in the biological context this observation corresponds to preparation toward differentiation. In Fig.~\ref{fig:peak_front}, we show that an isolated spot at $r=0$, in a circular domain, embedded initially in the stable homogeneous state $A_0$ initiates the formation of a circular propagating front that flips the background state from $A_0$ to $A_*$. After a further transient ($t\sim 3000$) the radially propagating front breaks up~\cite{yochelis2021nonlinear} triggering a new spot at the location of the front. The formation of this new spot destabilizes the original spot at $r=0$ which decays to $A_*$. Note that this behavior takes place in a parameter regime in which an isolated spot embedded in a $A_*$ background is normally stable. The nucleation of a new spot at the location of the front is of considerable interest at it may be associated with the process of side-branching~\cite{yochelis2021nonlinear,knobloch2021stationary}; our simulations indicate that such spots remain attached to the front and so continue to propagate at the front speed.

Some properties of spots supported by the spatial inhomogeneity associated with a propagating front have been discussed in Refs.~\cite{yochelis2021nonlinear,knobloch2021stationary}; a more detailed analysis of such fronts and their stability properties will be discussed elsewhere.

\section{Discussion}\label{sec:discussion}

 {Periodic spot patterns are known to emerge naturally in spatially extended reaction-diffusion models and in many cases the patterns self-organize in the vicinity of a Turing instability onset~\cite{pearson1993complex,kapral1995pattern,maini1997spatial}. However, reaction-diffusion spots can also coexist as isolated states, resembling particles embedded in a homogeneneous background~\cite{yochelis2008formation,brena2014subcritical}. The latter can be attributed to mutual repulsion between neighboring spots whenever they exhibit monotonic decay toward the rest state~\cite{knobloch2015spatial}. In Ref.~\cite{knobloch2021stationary} we have shown that in this case single- and multi-peak states in 1D are organized in an intricate bifurcation diagram referred to as foliated snaking. The foliated snaking scenario is generic but for applications, it is also necessary to understand the stability properties of the resulting states, and in particular, the DNS-based observation that in 1D all such states decay to the background state $A_*$.}

We have carried out a systematic linear stability analysis of localized peak and multi-peak states in 1D. Owing to the cumbersome form of the model~\eqref{eq:AI}, the bifurcation analysis relies on numerical continuation using the AUTO package~\cite{doedel2012auto} to compute the solution profiles and on numerical eigenvalue computation to determine stability. We demonstrated, via the solution of a linear eigenvalue problem, that the localized peaks solutions along the primary $N=1$ branch are all unstable, with the number of unstable modes gradually increasing with the number of peaks comprising the solution. Specifically, we showed that the bifurcating small amplitude solutions are once unstable with one real eigenvalue with additional eigenvalues becoming unstable at each subsequent fold. Near the right folds all unstable eigenvalues are real, but complex eigenvalues result from collisions of some of these eigenvalues along the solution branches above each fold. The resulting oscillatory modes represent the dominant instability mechanism along substantial portions of the foliated snaking structure of these states (Fig.~\ref{fig:fold_evalues_cont}). This behavior resembles that found earlier in a binary fluid convection problem\cite{beaume2018three} but differs markedly from the stability properties of standard homoclinic snaking in which the same pair of eigenvalues is responsible for the repeated gain and loss of stability~\cite{knobloch2015spatial}. 

 {In 2D we do not have a detailed bifurcation diagram organizing single-spot and multi-spot states, and neither do we have a detailed linear stability analysis of these states. The main reason is that the model~\eqref{eq:AI} is rather complicated and therefore not readily amenable to analytical investigation in contrast to two-variable AI systems with polynomial structure, such as the FitzHugh-Nagumo, Gray-Scott, and Lugiato-Lefever systems~\cite{or1998spot,kolokolnikov2006stability,doelman2009pulse,chen2011stability,jamieson2016localized,parra2018bifurcation,wong2021spot,verschueren2021dissecting}, as is already evident from the computation of the uniform states.} However, based on the DNS we can say with confidence that both single- and multi-spot states can be stable, and indeed are stable over a substantial range of values of the control parameter $\rH$. This conclusion is based on long time integration of multi-spot configurations with different boundary conditions. Our computations demonstrate the stability of these configurations in both amplitude and phase, and confirm the spot-spot repulsion expected on the basis of earlier results~\cite{knobloch2021stationary}. These results resemble those obtained in Ref.~\cite{chen2011stability} for the Gray-Scott model in the weak interaction regime in which spots interact only via their exponential tails, and in the semi-strong regime in which the dynamics of a particular spot is coupled to the instantaneous location of the other spots as well as the shape of the domain boundary via an explicitly constructed Green's function. In our multi-variable case, such an explicit construction is not possible and our spot computations span both distinguished regimes. 

In addition, we identified an intriguing and distinct transition in which a spot embedded in the competing but metastable homogeneous state $A_0$ emits a circular front which flips the background state from $A_0$ to $A_*$ and subsequently triggers the growth of a new spot at the location of the front. Transitions of this type require the coexistence of two simultaneously stable homogeneous states, and so are excluded in many simpler AI models. {The growth of this new spot destabilizes the original spot and leads to a localized state that ultimately propagates with a constant speed, implying not only that the side-branching model~\eqref{eq:AI} is able to capture robust spots but also its potential applicability to other chemical reactions involving inorganic developmental-like morphogenesis~\cite{malchow2019nonlinear}. On the other hand, in 1D our computations reveal that peaks located at a front connecting two stable homogeneous states may be stabilized by the presence of the front, albeit with a superposed large amplitude oscillation~\cite{yochelis2021nonlinear}. This oscillation manifests itself as a strong oscillation in both the spot amplitude and in the front propagation speed, a phenomenon that has been reported in the context of somite formation~\cite{baker2008mathematical}. Except for the role played by the front in catalyzing this {\it surfing} behavior the resulting state resembles the state of jumping oscillons studied in Refs.~\cite{yang2006jumping,cherkashin2008discontinuously,knobloch2021origin}.}

In summary, we have shown that while multi-variable activator-inhibitor (a.k.a. reaction-diffusion) models may bear some similarity to the two-variable models often used to scrutinize biological and chemical phenomena, they may also support new types of behavior that are absent from simpler models. Although multi-variable models inevitably complicate analytical approaches, insights derived from such approaches remain indispensable for understanding the more complex biological systems they seek to model. {We hope, therefore, that the results reported here may stimulate further studies of multi-variable reaction-diffusion models, not only in the context of modeling, but also to substantiate the role of new pattern formation mechanisms they support in two or more spatial dimensions.}

\begin{acknowledgments}
We have benefited from discussions with Nicol\'as Verschueren. This work was supported in part the National Science Foundation under grant DMS-1908891 (EK).
\end{acknowledgments}

\vskip 0.2in
\noindent The data that support the findings of this study are available from the corresponding author upon reasonable request.
\vskip 0.2in
\noindent \bf{REFERENCES}

\begin{thebibliography}{54}%
	\makeatletter
	\providecommand \@ifxundefined [1]{%
		\@ifx{#1\undefined}
	}%
	\providecommand \@ifnum [1]{%
		\ifnum #1\expandafter \@firstoftwo
		\else \expandafter \@secondoftwo
		\fi
	}%
	\providecommand \@ifx [1]{%
		\ifx #1\expandafter \@firstoftwo
		\else \expandafter \@secondoftwo
		\fi
	}%
	\providecommand \natexlab [1]{#1}%
	\providecommand \enquote  [1]{``#1''}%
	\providecommand \bibnamefont  [1]{#1}%
	\providecommand \bibfnamefont [1]{#1}%
	\providecommand \citenamefont [1]{#1}%
	\providecommand \href@noop [0]{\@secondoftwo}%
	\providecommand \href [0]{\begingroup \@sanitize@url \@href}%
	\providecommand \@href[1]{\@@startlink{#1}\@@href}%
	\providecommand \@@href[1]{\endgroup#1\@@endlink}%
	\providecommand \@sanitize@url [0]{\catcode `\\12\catcode `\$12\catcode
		`\&12\catcode `\#12\catcode `\^12\catcode `\_12\catcode `\%12\relax}%
	\providecommand \@@startlink[1]{}%
	\providecommand \@@endlink[0]{}%
	\providecommand \url  [0]{\begingroup\@sanitize@url \@url }%
	\providecommand \@url [1]{\endgroup\@href {#1}{\urlprefix }}%
	\providecommand \urlprefix  [0]{URL }%
	\providecommand \Eprint [0]{\href }%
	\providecommand \doibase [0]{https://doi.org/}%
	\providecommand \selectlanguage [0]{\@gobble}%
	\providecommand \bibinfo  [0]{\@secondoftwo}%
	\providecommand \bibfield  [0]{\@secondoftwo}%
	\providecommand \translation [1]{[#1]}%
	\providecommand \BibitemOpen [0]{}%
	\providecommand \bibitemStop [0]{}%
	\providecommand \bibitemNoStop [0]{.\EOS\space}%
	\providecommand \EOS [0]{\spacefactor3000\relax}%
	\providecommand \BibitemShut  [1]{\csname bibitem#1\endcsname}%
	\let\auto@bib@innerbib\@empty
	\bibitem [{\citenamefont {Turing}(1952)}]{tu52}%
	\BibitemOpen
	\bibfield  {author} {\bibinfo {author} {\bibfnamefont {A.}~\bibnamefont
			{Turing}},\ }\href@noop {} {\bibfield  {journal} {\bibinfo  {journal}
			{Philosophical Transactions of the Royal Society B}\ }\textbf {\bibinfo
			{volume} {237}},\ \bibinfo {pages} {37} (\bibinfo {year} {1952})}\BibitemShut
	{NoStop}%
	\bibitem [{\citenamefont {Cross}\ and\ \citenamefont {Hohenberg}(1993)}]{ch93}%
	\BibitemOpen
	\bibfield  {author} {\bibinfo {author} {\bibfnamefont {M.~C.}\ \bibnamefont
			{Cross}}\ and\ \bibinfo {author} {\bibfnamefont {P.~C.}\ \bibnamefont
			{Hohenberg}},\ }\href@noop {} {\bibfield  {journal} {\bibinfo  {journal}
			{Reviews of Modern Physics}\ }\textbf {\bibinfo {volume} {65}},\ \bibinfo
		{pages} {851} (\bibinfo {year} {1993})}\BibitemShut {NoStop}%
	\bibitem [{\citenamefont {Knobloch}(2015)}]{knobloch2015spatial}%
	\BibitemOpen
	\bibfield  {author} {\bibinfo {author} {\bibfnamefont {E.}~\bibnamefont
			{Knobloch}},\ }\href@noop {} {\bibfield  {journal} {\bibinfo  {journal}
			{Annual Review of Condensed Matter Physics}\ }\textbf {\bibinfo {volume}
			{6}},\ \bibinfo {pages} {325} (\bibinfo {year} {2015})}\BibitemShut {NoStop}%
	\bibitem [{\citenamefont {Burke}\ and\ \citenamefont
		{Knobloch}(2007)}]{burke2007snakes}%
	\BibitemOpen
	\bibfield  {author} {\bibinfo {author} {\bibfnamefont {J.}~\bibnamefont
			{Burke}}\ and\ \bibinfo {author} {\bibfnamefont {E.}~\bibnamefont
			{Knobloch}},\ }\href@noop {} {\bibfield  {journal} {\bibinfo  {journal}
			{Physics Letters A}\ }\textbf {\bibinfo {volume} {360}},\ \bibinfo {pages}
		{681} (\bibinfo {year} {2007})}\BibitemShut {NoStop}%
	\bibitem [{\citenamefont {Woods}\ and\ \citenamefont
		{Champneys}(1999)}]{woods1999heteroclinic}%
	\BibitemOpen
	\bibfield  {author} {\bibinfo {author} {\bibfnamefont {P.~D.}\ \bibnamefont
			{Woods}}\ and\ \bibinfo {author} {\bibfnamefont {A.~R.}\ \bibnamefont
			{Champneys}},\ }\href@noop {} {\bibfield  {journal} {\bibinfo  {journal}
			{Physica D}\ }\textbf {\bibinfo {volume} {129}},\ \bibinfo {pages} {147}
		(\bibinfo {year} {1999})}\BibitemShut {NoStop}%
	\bibitem [{\citenamefont {Knobloch}\ and\ \citenamefont
		{Yochelis}(2021)}]{knobloch2021stationary}%
	\BibitemOpen
	\bibfield  {author} {\bibinfo {author} {\bibfnamefont {E.}~\bibnamefont
			{Knobloch}}\ and\ \bibinfo {author} {\bibfnamefont {A.}~\bibnamefont
			{Yochelis}},\ }\href@noop {} {\bibfield  {journal} {\bibinfo  {journal} {IMA
				Journal of Applied Mathematics}\ }\textbf {\bibinfo {volume} {86}},\ \bibinfo
		{pages} {1066} (\bibinfo {year} {2021})}\BibitemShut {NoStop}%
	\bibitem [{\citenamefont {Meinhardt}(1976)}]{meinhardt1976morphogenesis}%
	\BibitemOpen
	\bibfield  {author} {\bibinfo {author} {\bibfnamefont {H.}~\bibnamefont
			{Meinhardt}},\ }\href@noop {} {\bibfield  {journal} {\bibinfo  {journal}
			{Differentiation}\ }\textbf {\bibinfo {volume} {6}},\ \bibinfo {pages} {117}
		(\bibinfo {year} {1976})}\BibitemShut {NoStop}%
	\bibitem [{\citenamefont {Yao}\ \emph {et~al.}(2007)\citenamefont {Yao},
		\citenamefont {Nowak}, \citenamefont {Yochelis}, \citenamefont {Garfinkel},\
		and\ \citenamefont {Bostr{\"o}m}}]{yao2007matrix}%
	\BibitemOpen
	\bibfield  {author} {\bibinfo {author} {\bibfnamefont {Y.}~\bibnamefont
			{Yao}}, \bibinfo {author} {\bibfnamefont {S.}~\bibnamefont {Nowak}}, \bibinfo
		{author} {\bibfnamefont {A.}~\bibnamefont {Yochelis}}, \bibinfo {author}
		{\bibfnamefont {A.}~\bibnamefont {Garfinkel}},\ and\ \bibinfo {author}
		{\bibfnamefont {K.~I.}\ \bibnamefont {Bostr{\"o}m}},\ }\href@noop {}
	{\bibfield  {journal} {\bibinfo  {journal} {Journal of Biological Chemistry}\
		}\textbf {\bibinfo {volume} {282}},\ \bibinfo {pages} {30131} (\bibinfo
		{year} {2007})}\BibitemShut {NoStop}%
	\bibitem [{\citenamefont {Metzger}\ \emph {et~al.}(2008)\citenamefont
		{Metzger}, \citenamefont {Klein}, \citenamefont {Martin},\ and\ \citenamefont
		{Krasnow}}]{metzger2008branching}%
	\BibitemOpen
	\bibfield  {author} {\bibinfo {author} {\bibfnamefont {R.~J.}\ \bibnamefont
			{Metzger}}, \bibinfo {author} {\bibfnamefont {O.~D.}\ \bibnamefont {Klein}},
		\bibinfo {author} {\bibfnamefont {G.~R.}\ \bibnamefont {Martin}},\ and\
		\bibinfo {author} {\bibfnamefont {M.~A.}\ \bibnamefont {Krasnow}},\
	}\href@noop {} {\bibfield  {journal} {\bibinfo  {journal} {Nature}\ }\textbf
		{\bibinfo {volume} {453}},\ \bibinfo {pages} {745} (\bibinfo {year}
		{2008})}\BibitemShut {NoStop}%
	\bibitem [{\citenamefont {Hirashima}\ \emph {et~al.}(2009)\citenamefont
		{Hirashima}, \citenamefont {Iwasa},\ and\ \citenamefont
		{Morishita}}]{hirashima2009mechanisms}%
	\BibitemOpen
	\bibfield  {author} {\bibinfo {author} {\bibfnamefont {T.}~\bibnamefont
			{Hirashima}}, \bibinfo {author} {\bibfnamefont {Y.}~\bibnamefont {Iwasa}},\
		and\ \bibinfo {author} {\bibfnamefont {Y.}~\bibnamefont {Morishita}},\
	}\href@noop {} {\bibfield  {journal} {\bibinfo  {journal} {Developmental
				Dynamics}\ }\textbf {\bibinfo {volume} {238}},\ \bibinfo {pages} {2813}
		(\bibinfo {year} {2009})}\BibitemShut {NoStop}%
	\bibitem [{\citenamefont {Menshykau}\ \emph {et~al.}(2012)\citenamefont
		{Menshykau}, \citenamefont {Kraemer},\ and\ \citenamefont
		{Iber}}]{menshykau2012branch}%
	\BibitemOpen
	\bibfield  {author} {\bibinfo {author} {\bibfnamefont {D.}~\bibnamefont
			{Menshykau}}, \bibinfo {author} {\bibfnamefont {C.}~\bibnamefont {Kraemer}},\
		and\ \bibinfo {author} {\bibfnamefont {D.}~\bibnamefont {Iber}},\ }\href@noop
	{} {\bibfield  {journal} {\bibinfo  {journal} {PLoS Computational Biology}\
		}\textbf {\bibinfo {volume} {8}} (\bibinfo {year} {2012})}\BibitemShut
	{NoStop}%
	\bibitem [{\citenamefont {Blanc}\ \emph {et~al.}(2012)\citenamefont {Blanc},
		\citenamefont {Coste}, \citenamefont {Pouchin}, \citenamefont {Aza{\"\i}s},
		\citenamefont {Blanchon}, \citenamefont {Gallot},\ and\ \citenamefont
		{Sapin}}]{blanc2012role}%
	\BibitemOpen
	\bibfield  {author} {\bibinfo {author} {\bibfnamefont {P.}~\bibnamefont
			{Blanc}}, \bibinfo {author} {\bibfnamefont {K.}~\bibnamefont {Coste}},
		\bibinfo {author} {\bibfnamefont {P.}~\bibnamefont {Pouchin}}, \bibinfo
		{author} {\bibfnamefont {J.-M.}\ \bibnamefont {Aza{\"\i}s}}, \bibinfo
		{author} {\bibfnamefont {L.}~\bibnamefont {Blanchon}}, \bibinfo {author}
		{\bibfnamefont {D.}~\bibnamefont {Gallot}},\ and\ \bibinfo {author}
		{\bibfnamefont {V.}~\bibnamefont {Sapin}},\ }\href@noop {} {\bibfield
		{journal} {\bibinfo  {journal} {PLoS One}\ }\textbf {\bibinfo {volume} {7}},\
		\bibinfo {pages} {e41643} (\bibinfo {year} {2012})}\BibitemShut {NoStop}%
	\bibitem [{\citenamefont {Celli{\`e}re}\ \emph {et~al.}(2012)\citenamefont
		{Celli{\`e}re}, \citenamefont {Menshykau},\ and\ \citenamefont
		{Iber}}]{celliere2012simulations}%
	\BibitemOpen
	\bibfield  {author} {\bibinfo {author} {\bibfnamefont {G.}~\bibnamefont
			{Celli{\`e}re}}, \bibinfo {author} {\bibfnamefont {D.}~\bibnamefont
			{Menshykau}},\ and\ \bibinfo {author} {\bibfnamefont {D.}~\bibnamefont
			{Iber}},\ }\href@noop {} {\bibfield  {journal} {\bibinfo  {journal} {Biology
				Open}\ }\textbf {\bibinfo {volume} {1}},\ \bibinfo {pages} {775} (\bibinfo
		{year} {2012})}\BibitemShut {NoStop}%
	\bibitem [{\citenamefont {Guo}\ \emph {et~al.}(2014{\natexlab{a}})\citenamefont
		{Guo}, \citenamefont {Chen}, \citenamefont {Zeng}, \citenamefont {Warburton},
		\citenamefont {Bostr{\"o}m}, \citenamefont {Ho}, \citenamefont {Zhao},\ and\
		\citenamefont {Garfinkel}}]{guo2014branching}%
	\BibitemOpen
	\bibfield  {author} {\bibinfo {author} {\bibfnamefont {Y.~A.}\ \bibnamefont
			{Guo}}, \bibinfo {author} {\bibfnamefont {T.-H.}\ \bibnamefont {Chen}},
		\bibinfo {author} {\bibfnamefont {X.}~\bibnamefont {Zeng}}, \bibinfo {author}
		{\bibfnamefont {D.}~\bibnamefont {Warburton}}, \bibinfo {author}
		{\bibfnamefont {K.~I.}\ \bibnamefont {Bostr{\"o}m}}, \bibinfo {author}
		{\bibfnamefont {C.-M.}\ \bibnamefont {Ho}}, \bibinfo {author} {\bibfnamefont
			{X.}~\bibnamefont {Zhao}},\ and\ \bibinfo {author} {\bibfnamefont
			{A.}~\bibnamefont {Garfinkel}},\ }\href@noop {} {\bibfield  {journal}
		{\bibinfo  {journal} {The Journal of Physiology}\ }\textbf {\bibinfo {volume}
			{592}},\ \bibinfo {pages} {313} (\bibinfo {year}
		{2014}{\natexlab{a}})}\BibitemShut {NoStop}%
	\bibitem [{\citenamefont {Guo}\ \emph {et~al.}(2014{\natexlab{b}})\citenamefont
		{Guo}, \citenamefont {Sun}, \citenamefont {Garfinkel},\ and\ \citenamefont
		{Zhao}}]{guo2014mechanisms}%
	\BibitemOpen
	\bibfield  {author} {\bibinfo {author} {\bibfnamefont {Y.~A.}\ \bibnamefont
			{Guo}}, \bibinfo {author} {\bibfnamefont {M.}~\bibnamefont {Sun}}, \bibinfo
		{author} {\bibfnamefont {A.}~\bibnamefont {Garfinkel}},\ and\ \bibinfo
		{author} {\bibfnamefont {X.}~\bibnamefont {Zhao}},\ }\href@noop {} {\bibfield
		{journal} {\bibinfo  {journal} {PloS One}\ }\textbf {\bibinfo {volume} {9}}
		(\bibinfo {year} {2014}{\natexlab{b}})}\BibitemShut {NoStop}%
	\bibitem [{\citenamefont {Hannezo}\ \emph {et~al.}(2017)\citenamefont
		{Hannezo}, \citenamefont {Scheele}, \citenamefont {Moad}, \citenamefont
		{Drogo}, \citenamefont {Heer}, \citenamefont {Sampogna}, \citenamefont
		{Van~Rheenen},\ and\ \citenamefont {Simons}}]{hannezo2017unifying}%
	\BibitemOpen
	\bibfield  {author} {\bibinfo {author} {\bibfnamefont {E.}~\bibnamefont
			{Hannezo}}, \bibinfo {author} {\bibfnamefont {C.~L. G.~J.}\ \bibnamefont
			{Scheele}}, \bibinfo {author} {\bibfnamefont {M.}~\bibnamefont {Moad}},
		\bibinfo {author} {\bibfnamefont {N.}~\bibnamefont {Drogo}}, \bibinfo
		{author} {\bibfnamefont {R.}~\bibnamefont {Heer}}, \bibinfo {author}
		{\bibfnamefont {R.~V.}\ \bibnamefont {Sampogna}}, \bibinfo {author}
		{\bibfnamefont {J.}~\bibnamefont {Van~Rheenen}},\ and\ \bibinfo {author}
		{\bibfnamefont {B.~D.}\ \bibnamefont {Simons}},\ }\href@noop {} {\bibfield
		{journal} {\bibinfo  {journal} {Cell}\ }\textbf {\bibinfo {volume} {171}},\
		\bibinfo {pages} {242} (\bibinfo {year} {2017})}\BibitemShut {NoStop}%
	\bibitem [{\citenamefont {Xu}\ \emph {et~al.}(2017)\citenamefont {Xu},
		\citenamefont {Sun},\ and\ \citenamefont {Zhao}}]{xu2017turing}%
	\BibitemOpen
	\bibfield  {author} {\bibinfo {author} {\bibfnamefont {H.}~\bibnamefont
			{Xu}}, \bibinfo {author} {\bibfnamefont {M.}~\bibnamefont {Sun}},\ and\
		\bibinfo {author} {\bibfnamefont {X.}~\bibnamefont {Zhao}},\ }\href@noop {}
	{\bibfield  {journal} {\bibinfo  {journal} {PloS One}\ }\textbf {\bibinfo
			{volume} {12}} (\bibinfo {year} {2017})}\BibitemShut {NoStop}%
	\bibitem [{\citenamefont {Shan}\ \emph {et~al.}(2018)\citenamefont {Shan},
		\citenamefont {Chuan-shan}, \citenamefont {Ming-zhu},\ and\ \citenamefont
		{Xin}}]{shan2018meshwork}%
	\BibitemOpen
	\bibfield  {author} {\bibinfo {author} {\bibfnamefont {G.}~\bibnamefont
			{Shan}}, \bibinfo {author} {\bibfnamefont {H.}~\bibnamefont {Chuan-shan}},
		\bibinfo {author} {\bibfnamefont {S.}~\bibnamefont {Ming-zhu}},\ and\
		\bibinfo {author} {\bibfnamefont {Z.}~\bibnamefont {Xin}},\ }\href@noop {}
	{\bibfield  {journal} {\bibinfo  {journal} {Journal of Theoretical Biology}\
		}\textbf {\bibinfo {volume} {455}},\ \bibinfo {pages} {293} (\bibinfo {year}
		{2018})}\BibitemShut {NoStop}%
	\bibitem [{\citenamefont {Zhu}\ and\ \citenamefont
		{Yang}(2018)}]{zhu2018turing}%
	\BibitemOpen
	\bibfield  {author} {\bibinfo {author} {\bibfnamefont {X.}~\bibnamefont
			{Zhu}}\ and\ \bibinfo {author} {\bibfnamefont {H.}~\bibnamefont {Yang}},\
	}\href@noop {} {\bibfield  {journal} {\bibinfo  {journal} {Micromachines}\
		}\textbf {\bibinfo {volume} {9}},\ \bibinfo {pages} {109} (\bibinfo {year}
		{2018})}\BibitemShut {NoStop}%
	\bibitem [{\citenamefont {Sainio}\ \emph {et~al.}(1997)\citenamefont {Sainio},
		\citenamefont {Suvanto}, \citenamefont {Davies}, \citenamefont {Wartiovaara},
		\citenamefont {Wartiovaara}, \citenamefont {Saarma}, \citenamefont {Arumae},
		\citenamefont {Meng}, \citenamefont {Lindahl}, \citenamefont {Pachnis} \emph
		{et~al.}}]{sainio1997glial}%
	\BibitemOpen
	\bibfield  {author} {\bibinfo {author} {\bibfnamefont {K.}~\bibnamefont
			{Sainio}}, \bibinfo {author} {\bibfnamefont {P.}~\bibnamefont {Suvanto}},
		\bibinfo {author} {\bibfnamefont {J.}~\bibnamefont {Davies}}, \bibinfo
		{author} {\bibfnamefont {J.}~\bibnamefont {Wartiovaara}}, \bibinfo {author}
		{\bibfnamefont {K.}~\bibnamefont {Wartiovaara}}, \bibinfo {author}
		{\bibfnamefont {M.}~\bibnamefont {Saarma}}, \bibinfo {author} {\bibfnamefont
			{U.}~\bibnamefont {Arumae}}, \bibinfo {author} {\bibfnamefont
			{X.}~\bibnamefont {Meng}}, \bibinfo {author} {\bibfnamefont {M.}~\bibnamefont
			{Lindahl}}, \bibinfo {author} {\bibfnamefont {V.}~\bibnamefont {Pachnis}},
		\emph {et~al.},\ }\href@noop {} {\bibfield  {journal} {\bibinfo  {journal}
			{Development}\ }\textbf {\bibinfo {volume} {124}},\ \bibinfo {pages} {4077}
		(\bibinfo {year} {1997})}\BibitemShut {NoStop}%
	\bibitem [{\citenamefont {Tang}\ \emph {et~al.}(1998)\citenamefont {Tang},
		\citenamefont {Worley}, \citenamefont {Sanicola},\ and\ \citenamefont
		{Dressler}}]{tang1998ret}%
	\BibitemOpen
	\bibfield  {author} {\bibinfo {author} {\bibfnamefont {M.-J.}\ \bibnamefont
			{Tang}}, \bibinfo {author} {\bibfnamefont {D.}~\bibnamefont {Worley}},
		\bibinfo {author} {\bibfnamefont {M.}~\bibnamefont {Sanicola}},\ and\
		\bibinfo {author} {\bibfnamefont {G.~R.}\ \bibnamefont {Dressler}},\
	}\href@noop {} {\bibfield  {journal} {\bibinfo  {journal} {The Journal of
				Cell Biology}\ }\textbf {\bibinfo {volume} {142}},\ \bibinfo {pages} {1337}
		(\bibinfo {year} {1998})}\BibitemShut {NoStop}%
	\bibitem [{\citenamefont {Lebeche}\ \emph {et~al.}(1999)\citenamefont
		{Lebeche}, \citenamefont {Malpel},\ and\ \citenamefont
		{Cardoso}}]{lebeche1999fibroblast}%
	\BibitemOpen
	\bibfield  {author} {\bibinfo {author} {\bibfnamefont {D.}~\bibnamefont
			{Lebeche}}, \bibinfo {author} {\bibfnamefont {S.}~\bibnamefont {Malpel}},\
		and\ \bibinfo {author} {\bibfnamefont {W.~V.}\ \bibnamefont {Cardoso}},\
	}\href@noop {} {\bibfield  {journal} {\bibinfo  {journal} {Mechanisms of
				Development}\ }\textbf {\bibinfo {volume} {86}},\ \bibinfo {pages} {125}
		(\bibinfo {year} {1999})}\BibitemShut {NoStop}%
	\bibitem [{\citenamefont {Tang}\ \emph {et~al.}(2002)\citenamefont {Tang},
		\citenamefont {Cai}, \citenamefont {Tsai}, \citenamefont {Wang},\ and\
		\citenamefont {Dressler}}]{tang2002ureteric}%
	\BibitemOpen
	\bibfield  {author} {\bibinfo {author} {\bibfnamefont {M.-J.}\ \bibnamefont
			{Tang}}, \bibinfo {author} {\bibfnamefont {Y.}~\bibnamefont {Cai}}, \bibinfo
		{author} {\bibfnamefont {S.-J.}\ \bibnamefont {Tsai}}, \bibinfo {author}
		{\bibfnamefont {Y.-K.}\ \bibnamefont {Wang}},\ and\ \bibinfo {author}
		{\bibfnamefont {G.~R.}\ \bibnamefont {Dressler}},\ }\href@noop {} {\bibfield
		{journal} {\bibinfo  {journal} {Developmental Biology}\ }\textbf {\bibinfo
			{volume} {243}},\ \bibinfo {pages} {128} (\bibinfo {year}
		{2002})}\BibitemShut {NoStop}%
	\bibitem [{\citenamefont {Gilbert}\ and\ \citenamefont
		{Rannels}(2004)}]{gilbert2004matrix}%
	\BibitemOpen
	\bibfield  {author} {\bibinfo {author} {\bibfnamefont {K.~A.}\ \bibnamefont
			{Gilbert}}\ and\ \bibinfo {author} {\bibfnamefont {S.~R.}\ \bibnamefont
			{Rannels}},\ }\href@noop {} {\bibfield  {journal} {\bibinfo  {journal}
			{American Journal of Physiology-Lung Cellular and Molecular Physiology}\
		}\textbf {\bibinfo {volume} {286}},\ \bibinfo {pages} {L1179} (\bibinfo
		{year} {2004})}\BibitemShut {NoStop}%
	\bibitem [{\citenamefont {Affolter}\ \emph {et~al.}(2009)\citenamefont
		{Affolter}, \citenamefont {Zeller},\ and\ \citenamefont
		{Caussinus}}]{affolter2009tissue}%
	\BibitemOpen
	\bibfield  {author} {\bibinfo {author} {\bibfnamefont {M.}~\bibnamefont
			{Affolter}}, \bibinfo {author} {\bibfnamefont {R.}~\bibnamefont {Zeller}},\
		and\ \bibinfo {author} {\bibfnamefont {E.}~\bibnamefont {Caussinus}},\
	}\href@noop {} {\bibfield  {journal} {\bibinfo  {journal} {Nature Reviews
				Molecular Cell Biology}\ }\textbf {\bibinfo {volume} {10}},\ \bibinfo {pages}
		{831} (\bibinfo {year} {2009})}\BibitemShut {NoStop}%
	\bibitem [{\citenamefont {Yao}\ \emph {et~al.}(2011)\citenamefont {Yao},
		\citenamefont {Jumabay}, \citenamefont {Wang},\ and\ \citenamefont
		{Bostr{\"o}m}}]{yao2011matrix}%
	\BibitemOpen
	\bibfield  {author} {\bibinfo {author} {\bibfnamefont {Y.}~\bibnamefont
			{Yao}}, \bibinfo {author} {\bibfnamefont {M.}~\bibnamefont {Jumabay}},
		\bibinfo {author} {\bibfnamefont {A.}~\bibnamefont {Wang}},\ and\ \bibinfo
		{author} {\bibfnamefont {K.~I.}\ \bibnamefont {Bostr{\"o}m}},\ }\href@noop {}
	{\bibfield  {journal} {\bibinfo  {journal} {The Journal of Clinical
				Investigation}\ }\textbf {\bibinfo {volume} {121}},\ \bibinfo {pages} {2993}
		(\bibinfo {year} {2011})}\BibitemShut {NoStop}%
	\bibitem [{\citenamefont {Hagiwara}\ \emph {et~al.}(2015)\citenamefont
		{Hagiwara}, \citenamefont {Peng},\ and\ \citenamefont
		{Ho}}]{hagiwara2015vitro}%
	\BibitemOpen
	\bibfield  {author} {\bibinfo {author} {\bibfnamefont {M.}~\bibnamefont
			{Hagiwara}}, \bibinfo {author} {\bibfnamefont {F.}~\bibnamefont {Peng}},\
		and\ \bibinfo {author} {\bibfnamefont {C.-M.}\ \bibnamefont {Ho}},\
	}\href@noop {} {\bibfield  {journal} {\bibinfo  {journal} {Scientific
				Reports}\ }\textbf {\bibinfo {volume} {5}},\ \bibinfo {pages} {8054}
		(\bibinfo {year} {2015})}\BibitemShut {NoStop}%
	\bibitem [{\citenamefont {Menshykau}\ \emph {et~al.}(2019)\citenamefont
		{Menshykau}, \citenamefont {Michos}, \citenamefont {Lang}, \citenamefont
		{Conrad}, \citenamefont {McMahon},\ and\ \citenamefont
		{Iber}}]{menshykau2019image}%
	\BibitemOpen
	\bibfield  {author} {\bibinfo {author} {\bibfnamefont {D.}~\bibnamefont
			{Menshykau}}, \bibinfo {author} {\bibfnamefont {O.}~\bibnamefont {Michos}},
		\bibinfo {author} {\bibfnamefont {C.}~\bibnamefont {Lang}}, \bibinfo {author}
		{\bibfnamefont {L.}~\bibnamefont {Conrad}}, \bibinfo {author} {\bibfnamefont
			{A.~P.}\ \bibnamefont {McMahon}},\ and\ \bibinfo {author} {\bibfnamefont
			{D.}~\bibnamefont {Iber}},\ }\href@noop {} {\bibfield  {journal} {\bibinfo
			{journal} {Nature Communications}\ }\textbf {\bibinfo {volume} {10}},\
		\bibinfo {pages} {1} (\bibinfo {year} {2019})}\BibitemShut {NoStop}%
	\bibitem [{\citenamefont {Yochelis}(2021)}]{yochelis2021nonlinear}%
	\BibitemOpen
	\bibfield  {author} {\bibinfo {author} {\bibfnamefont {A.}~\bibnamefont
			{Yochelis}},\ }\href@noop {} {\bibfield  {journal} {\bibinfo  {journal}
			{Chaos}\ }\textbf {\bibinfo {volume} {31}},\ \bibinfo {pages} {051102}
		(\bibinfo {year} {2021})}\BibitemShut {NoStop}%
	\bibitem [{\citenamefont {Yochelis}\ and\ \citenamefont
		{Garfinkel}(2008)}]{yochelis2008front}%
	\BibitemOpen
	\bibfield  {author} {\bibinfo {author} {\bibfnamefont {A.}~\bibnamefont
			{Yochelis}}\ and\ \bibinfo {author} {\bibfnamefont {A.}~\bibnamefont
			{Garfinkel}},\ }\href@noop {} {\bibfield  {journal} {\bibinfo  {journal}
			{Physical Review E}\ }\textbf {\bibinfo {volume} {77}},\ \bibinfo {pages}
		{035204} (\bibinfo {year} {2008})}\BibitemShut {NoStop}%
	\bibitem [{\citenamefont {Doedel}\ \emph {et~al.}(2012)\citenamefont {Doedel},
		\citenamefont {Champneys}, \citenamefont {Fairgrieve}, \citenamefont
		{Kuznetsov}, \citenamefont {Oldeman}, \citenamefont {Paffenroth},
		\citenamefont {Sandstede}, \citenamefont {Wang},\ and\ \citenamefont
		{Zhang}}]{doedel2012auto}%
	\BibitemOpen
	\bibfield  {author} {\bibinfo {author} {\bibfnamefont {E.~J.}\ \bibnamefont
			{Doedel}}, \bibinfo {author} {\bibfnamefont {A.~R.}\ \bibnamefont
			{Champneys}}, \bibinfo {author} {\bibfnamefont {T.}~\bibnamefont
			{Fairgrieve}}, \bibinfo {author} {\bibfnamefont {Y.}~\bibnamefont
			{Kuznetsov}}, \bibinfo {author} {\bibfnamefont {B.}~\bibnamefont {Oldeman}},
		\bibinfo {author} {\bibfnamefont {R.}~\bibnamefont {Paffenroth}}, \bibinfo
		{author} {\bibfnamefont {B.}~\bibnamefont {Sandstede}}, \bibinfo {author}
		{\bibfnamefont {X.}~\bibnamefont {Wang}},\ and\ \bibinfo {author}
		{\bibfnamefont {C.}~\bibnamefont {Zhang}},\ }\href@noop {} {\bibfield
		{journal} {\bibinfo  {journal} {{AUTO}07p: {C}ontinuation and bifurcation
				software for ordinary differential equations, Concordia University,
				http://indy.cs.concordia.ca/auto}\ } (\bibinfo {year} {2012})}\BibitemShut
	{NoStop}%
	\bibitem [{\citenamefont {Burke}\ and\ \citenamefont
		{Knobloch}(2006)}]{burke2006localized}%
	\BibitemOpen
	\bibfield  {author} {\bibinfo {author} {\bibfnamefont {J.}~\bibnamefont
			{Burke}}\ and\ \bibinfo {author} {\bibfnamefont {E.}~\bibnamefont
			{Knobloch}},\ }\href@noop {} {\bibfield  {journal} {\bibinfo  {journal}
			{Physical Review E}\ }\textbf {\bibinfo {volume} {73}},\ \bibinfo {pages}
		{056211} (\bibinfo {year} {2006})}\BibitemShut {NoStop}%
	\bibitem [{\citenamefont {Beaume}\ \emph {et~al.}(2018)\citenamefont {Beaume},
		\citenamefont {Bergeon},\ and\ \citenamefont {Knobloch}}]{beaume2018three}%
	\BibitemOpen
	\bibfield  {author} {\bibinfo {author} {\bibfnamefont {C.}~\bibnamefont
			{Beaume}}, \bibinfo {author} {\bibfnamefont {A.}~\bibnamefont {Bergeon}},\
		and\ \bibinfo {author} {\bibfnamefont {E.}~\bibnamefont {Knobloch}},\
	}\href@noop {} {\bibfield  {journal} {\bibinfo  {journal} {Journal of Fluid
				Mechanics}\ }\textbf {\bibinfo {volume} {840}},\ \bibinfo {pages} {74}
		(\bibinfo {year} {2018})}\BibitemShut {NoStop}%
	\bibitem [{\citenamefont {Kolokolnikov}\ \emph
		{et~al.}(2006{\natexlab{a}})\citenamefont {Kolokolnikov}, \citenamefont
		{Sun}, \citenamefont {Ward},\ and\ \citenamefont
		{Wei}}]{kolokolnikov2006stability}%
	\BibitemOpen
	\bibfield  {author} {\bibinfo {author} {\bibfnamefont {T.}~\bibnamefont
			{Kolokolnikov}}, \bibinfo {author} {\bibfnamefont {W.}~\bibnamefont {Sun}},
		\bibinfo {author} {\bibfnamefont {M.}~\bibnamefont {Ward}},\ and\ \bibinfo
		{author} {\bibfnamefont {J.}~\bibnamefont {Wei}},\ }\href@noop {} {\bibfield
		{journal} {\bibinfo  {journal} {SIAM Journal on Applied Dynamical Systems}\
		}\textbf {\bibinfo {volume} {5}},\ \bibinfo {pages} {313} (\bibinfo {year}
		{2006}{\natexlab{a}})}\BibitemShut {NoStop}%
	\bibitem [{\citenamefont {Kolokolnikov}\ \emph
		{et~al.}(2006{\natexlab{b}})\citenamefont {Kolokolnikov}, \citenamefont
		{Ward},\ and\ \citenamefont {Wei}}]{kolokolnikov2006zigzag}%
	\BibitemOpen
	\bibfield  {author} {\bibinfo {author} {\bibfnamefont {T.}~\bibnamefont
			{Kolokolnikov}}, \bibinfo {author} {\bibfnamefont {M.~J.}\ \bibnamefont
			{Ward}},\ and\ \bibinfo {author} {\bibfnamefont {J.}~\bibnamefont {Wei}},\
	}\href@noop {} {\bibfield  {journal} {\bibinfo  {journal} {Studies in Applied
				Mathematics}\ }\textbf {\bibinfo {volume} {116}},\ \bibinfo {pages} {35}
		(\bibinfo {year} {2006}{\natexlab{b}})}\BibitemShut {NoStop}%
	\bibitem [{\citenamefont {Yochelis}\ \emph {et~al.}(2008)\citenamefont
		{Yochelis}, \citenamefont {Tintut}, \citenamefont {Demer},\ and\
		\citenamefont {Garfinkel}}]{yochelis2008formation}%
	\BibitemOpen
	\bibfield  {author} {\bibinfo {author} {\bibfnamefont {A.}~\bibnamefont
			{Yochelis}}, \bibinfo {author} {\bibfnamefont {Y.}~\bibnamefont {Tintut}},
		\bibinfo {author} {\bibfnamefont {L.~L.}\ \bibnamefont {Demer}},\ and\
		\bibinfo {author} {\bibfnamefont {A.}~\bibnamefont {Garfinkel}},\ }\href@noop
	{} {\bibfield  {journal} {\bibinfo  {journal} {New Journal of Physics}\
		}\textbf {\bibinfo {volume} {10}},\ \bibinfo {pages} {055002} (\bibinfo
		{year} {2008})}\BibitemShut {NoStop}%
	\bibitem [{\citenamefont {Pearson}(1993)}]{pearson1993complex}%
	\BibitemOpen
	\bibfield  {author} {\bibinfo {author} {\bibfnamefont {J.~E.}\ \bibnamefont
			{Pearson}},\ }\href@noop {} {\bibfield  {journal} {\bibinfo  {journal}
			{Science}\ }\textbf {\bibinfo {volume} {261}},\ \bibinfo {pages} {189}
		(\bibinfo {year} {1993})}\BibitemShut {NoStop}%
	\bibitem [{\citenamefont {Purwins}\ \emph {et~al.}(2005)\citenamefont
		{Purwins}, \citenamefont {B{\"o}deker},\ and\ \citenamefont
		{Liehr}}]{purwins2005dissipative}%
	\BibitemOpen
	\bibfield  {author} {\bibinfo {author} {\bibfnamefont {H.-G.}\ \bibnamefont
			{Purwins}}, \bibinfo {author} {\bibfnamefont {H.}~\bibnamefont
			{B{\"o}deker}},\ and\ \bibinfo {author} {\bibfnamefont {A.}~\bibnamefont
			{Liehr}},\ }in\ \href@noop {} {\emph {\bibinfo {booktitle} {Dissipative
				Solitons}}}\ (\bibinfo  {publisher} {Springer},\ \bibinfo {year} {2005})\
	pp.\ \bibinfo {pages} {267--308}\BibitemShut {NoStop}%
	\bibitem [{\citenamefont {Vanag}\ and\ \citenamefont
		{Epstein}(2007)}]{vanag2007localized}%
	\BibitemOpen
	\bibfield  {author} {\bibinfo {author} {\bibfnamefont {V.~K.}\ \bibnamefont
			{Vanag}}\ and\ \bibinfo {author} {\bibfnamefont {I.~R.}\ \bibnamefont
			{Epstein}},\ }\href@noop {} {\bibfield  {journal} {\bibinfo  {journal}
			{Chaos}\ }\textbf {\bibinfo {volume} {17}},\ \bibinfo {pages} {037110}
		(\bibinfo {year} {2007})}\BibitemShut {NoStop}%
	\bibitem [{\citenamefont {Chen}\ and\ \citenamefont
		{Ward}(2011)}]{chen2011stability}%
	\BibitemOpen
	\bibfield  {author} {\bibinfo {author} {\bibfnamefont {W.}~\bibnamefont
			{Chen}}\ and\ \bibinfo {author} {\bibfnamefont {M.~J.}\ \bibnamefont
			{Ward}},\ }\href@noop {} {\bibfield  {journal} {\bibinfo  {journal} {SIAM
				Journal on Applied Dynamical Systems}\ }\textbf {\bibinfo {volume} {10}},\
		\bibinfo {pages} {582} (\bibinfo {year} {2011})}\BibitemShut {NoStop}%
	\bibitem [{\citenamefont {Jamieson-Lane}\ \emph {et~al.}(2016)\citenamefont
		{Jamieson-Lane}, \citenamefont {Trinh},\ and\ \citenamefont
		{Ward}}]{jamieson2016localized}%
	\BibitemOpen
	\bibfield  {author} {\bibinfo {author} {\bibfnamefont {A.}~\bibnamefont
			{Jamieson-Lane}}, \bibinfo {author} {\bibfnamefont {P.~H.}\ \bibnamefont
			{Trinh}},\ and\ \bibinfo {author} {\bibfnamefont {M.~J.}\ \bibnamefont
			{Ward}},\ }in\ \href@noop {} {\emph {\bibinfo {booktitle} {Mathematical and
				Computational Approaches in Advancing Modern Science and Engineering}}}\
	(\bibinfo  {publisher} {Springer},\ \bibinfo {year} {2016})\ pp.\ \bibinfo
	{pages} {641--651}\BibitemShut {NoStop}%
	\bibitem [{\citenamefont {Wong}\ and\ \citenamefont
		{Ward}(2021)}]{wong2021spot}%
	\BibitemOpen
	\bibfield  {author} {\bibinfo {author} {\bibfnamefont {T.}~\bibnamefont
			{Wong}}\ and\ \bibinfo {author} {\bibfnamefont {M.~J.}\ \bibnamefont
			{Ward}},\ }\href@noop {} {\bibfield  {journal} {\bibinfo  {journal} {Studies
				in Applied Mathematics}\ }\textbf {\bibinfo {volume} {146}},\ \bibinfo
		{pages} {779} (\bibinfo {year} {2021})}\BibitemShut {NoStop}%
	\bibitem [{\citenamefont {Kapral}(1995)}]{kapral1995pattern}%
	\BibitemOpen
	\bibfield  {author} {\bibinfo {author} {\bibfnamefont {R.}~\bibnamefont
			{Kapral}},\ }\href@noop {} {\bibfield  {journal} {\bibinfo  {journal}
			{Physica D}\ }\textbf {\bibinfo {volume} {86}},\ \bibinfo {pages} {149}
		(\bibinfo {year} {1995})}\BibitemShut {NoStop}%
	\bibitem [{\citenamefont {Maini}\ \emph {et~al.}(1997)\citenamefont {Maini},
		\citenamefont {Painter},\ and\ \citenamefont {Chau}}]{maini1997spatial}%
	\BibitemOpen
	\bibfield  {author} {\bibinfo {author} {\bibfnamefont {P.}~\bibnamefont
			{Maini}}, \bibinfo {author} {\bibfnamefont {K.}~\bibnamefont {Painter}},\
		and\ \bibinfo {author} {\bibfnamefont {H.~P.}\ \bibnamefont {Chau}},\
	}\href@noop {} {\bibfield  {journal} {\bibinfo  {journal} {Journal of the
				Chemical Society, Faraday Transactions}\ }\textbf {\bibinfo {volume} {93}},\
		\bibinfo {pages} {3601} (\bibinfo {year} {1997})}\BibitemShut {NoStop}%
	\bibitem [{\citenamefont {Bre{\~n}a-Medina}\ and\ \citenamefont
		{Champneys}(2014)}]{brena2014subcritical}%
	\BibitemOpen
	\bibfield  {author} {\bibinfo {author} {\bibfnamefont {V.}~\bibnamefont
			{Bre{\~n}a-Medina}}\ and\ \bibinfo {author} {\bibfnamefont {A.}~\bibnamefont
			{Champneys}},\ }\href@noop {} {\bibfield  {journal} {\bibinfo  {journal}
			{Physical Review E}\ }\textbf {\bibinfo {volume} {90}},\ \bibinfo {pages}
		{032923} (\bibinfo {year} {2014})}\BibitemShut {NoStop}%
	\bibitem [{\citenamefont {Or-Guil}\ \emph {et~al.}(1998)\citenamefont
		{Or-Guil}, \citenamefont {Bode}, \citenamefont {Schenk},\ and\ \citenamefont
		{Purwins}}]{or1998spot}%
	\BibitemOpen
	\bibfield  {author} {\bibinfo {author} {\bibfnamefont {M.}~\bibnamefont
			{Or-Guil}}, \bibinfo {author} {\bibfnamefont {M.}~\bibnamefont {Bode}},
		\bibinfo {author} {\bibfnamefont {C.}~\bibnamefont {Schenk}},\ and\ \bibinfo
		{author} {\bibfnamefont {H.-G.}\ \bibnamefont {Purwins}},\ }\href@noop {}
	{\bibfield  {journal} {\bibinfo  {journal} {Physical Review E}\ }\textbf
		{\bibinfo {volume} {57}},\ \bibinfo {pages} {6432} (\bibinfo {year}
		{1998})}\BibitemShut {NoStop}%
	\bibitem [{\citenamefont {Doelman}\ \emph {et~al.}(2009)\citenamefont
		{Doelman}, \citenamefont {Van~Heijster},\ and\ \citenamefont
		{Kaper}}]{doelman2009pulse}%
	\BibitemOpen
	\bibfield  {author} {\bibinfo {author} {\bibfnamefont {A.}~\bibnamefont
			{Doelman}}, \bibinfo {author} {\bibfnamefont {P.}~\bibnamefont
			{Van~Heijster}},\ and\ \bibinfo {author} {\bibfnamefont {T.~J.}\ \bibnamefont
			{Kaper}},\ }\href@noop {} {\bibfield  {journal} {\bibinfo  {journal} {Journal
				of Dynamics and Differential Equations}\ }\textbf {\bibinfo {volume} {21}},\
		\bibinfo {pages} {73} (\bibinfo {year} {2009})}\BibitemShut {NoStop}%
	\bibitem [{\citenamefont {Parra-Rivas}\ \emph {et~al.}(2018)\citenamefont
		{Parra-Rivas}, \citenamefont {Gomila}, \citenamefont {Gelens},\ and\
		\citenamefont {Knobloch}}]{parra2018bifurcation}%
	\BibitemOpen
	\bibfield  {author} {\bibinfo {author} {\bibfnamefont {P.}~\bibnamefont
			{Parra-Rivas}}, \bibinfo {author} {\bibfnamefont {D.}~\bibnamefont {Gomila}},
		\bibinfo {author} {\bibfnamefont {L.}~\bibnamefont {Gelens}},\ and\ \bibinfo
		{author} {\bibfnamefont {E.}~\bibnamefont {Knobloch}},\ }\href@noop {}
	{\bibfield  {journal} {\bibinfo  {journal} {Physical Review E}\ }\textbf
		{\bibinfo {volume} {97}},\ \bibinfo {pages} {042204} (\bibinfo {year}
		{2018})}\BibitemShut {NoStop}%
	\bibitem [{\citenamefont {Verschueren}\ and\ \citenamefont
		{Champneys}(2021)}]{verschueren2021dissecting}%
	\BibitemOpen
	\bibfield  {author} {\bibinfo {author} {\bibfnamefont {N.}~\bibnamefont
			{Verschueren}}\ and\ \bibinfo {author} {\bibfnamefont {A.~R.}\ \bibnamefont
			{Champneys}},\ }\href@noop {} {\bibfield  {journal} {\bibinfo  {journal}
			{Physica D}\ }\textbf {\bibinfo {volume} {419}},\ \bibinfo {pages} {132858}
		(\bibinfo {year} {2021})}\BibitemShut {NoStop}%
	\bibitem [{\citenamefont {Malchow}\ \emph {et~al.}(2019)\citenamefont
		{Malchow}, \citenamefont {Azhand}, \citenamefont {Knoll}, \citenamefont
		{Engel},\ and\ \citenamefont {Steinbock}}]{malchow2019nonlinear}%
	\BibitemOpen
	\bibfield  {author} {\bibinfo {author} {\bibfnamefont {A.-K.}\ \bibnamefont
			{Malchow}}, \bibinfo {author} {\bibfnamefont {A.}~\bibnamefont {Azhand}},
		\bibinfo {author} {\bibfnamefont {P.}~\bibnamefont {Knoll}}, \bibinfo
		{author} {\bibfnamefont {H.}~\bibnamefont {Engel}},\ and\ \bibinfo {author}
		{\bibfnamefont {O.}~\bibnamefont {Steinbock}},\ }\href@noop {} {\bibfield
		{journal} {\bibinfo  {journal} {Chaos}\ }\textbf {\bibinfo {volume} {29}},\
		\bibinfo {pages} {053129} (\bibinfo {year} {2019})}\BibitemShut {NoStop}%
	\bibitem [{\citenamefont {Baker}\ \emph {et~al.}(2008)\citenamefont {Baker},
		\citenamefont {Schnell},\ and\ \citenamefont
		{Maini}}]{baker2008mathematical}%
	\BibitemOpen
	\bibfield  {author} {\bibinfo {author} {\bibfnamefont {R.~E.}\ \bibnamefont
			{Baker}}, \bibinfo {author} {\bibfnamefont {S.}~\bibnamefont {Schnell}},\
		and\ \bibinfo {author} {\bibfnamefont {P.~K.}\ \bibnamefont {Maini}},\
	}\href@noop {} {\bibfield  {journal} {\bibinfo  {journal} {Current Topics in
				Developmental Biology}\ }\textbf {\bibinfo {volume} {81}},\ \bibinfo {pages}
		{183} (\bibinfo {year} {2008})}\BibitemShut {NoStop}%
	\bibitem [{\citenamefont {Yang}\ \emph {et~al.}(2006)\citenamefont {Yang},
		\citenamefont {Zhabotinsky},\ and\ \citenamefont
		{Epstein}}]{yang2006jumping}%
	\BibitemOpen
	\bibfield  {author} {\bibinfo {author} {\bibfnamefont {L.}~\bibnamefont
			{Yang}}, \bibinfo {author} {\bibfnamefont {A.~M.}\ \bibnamefont
			{Zhabotinsky}},\ and\ \bibinfo {author} {\bibfnamefont {I.~R.}\ \bibnamefont
			{Epstein}},\ }\href@noop {} {\bibfield  {journal} {\bibinfo  {journal}
			{Physical Chemistry Chemical Physics}\ }\textbf {\bibinfo {volume} {8}},\
		\bibinfo {pages} {4647} (\bibinfo {year} {2006})}\BibitemShut {NoStop}%
	\bibitem [{\citenamefont {Cherkashin}\ \emph {et~al.}(2008)\citenamefont
		{Cherkashin}, \citenamefont {Vanag},\ and\ \citenamefont
		{Epstein}}]{cherkashin2008discontinuously}%
	\BibitemOpen
	\bibfield  {author} {\bibinfo {author} {\bibfnamefont {A.~A.}\ \bibnamefont
			{Cherkashin}}, \bibinfo {author} {\bibfnamefont {V.~K.}\ \bibnamefont
			{Vanag}},\ and\ \bibinfo {author} {\bibfnamefont {I.~R.}\ \bibnamefont
			{Epstein}},\ }\href@noop {} {\bibfield  {journal} {\bibinfo  {journal}
			{Journal of Chemical Physics}\ }\textbf {\bibinfo {volume} {128}},\ \bibinfo
		{pages} {204508} (\bibinfo {year} {2008})}\BibitemShut {NoStop}%
	\bibitem [{\citenamefont {Knobloch}\ \emph {et~al.}(2021)\citenamefont
		{Knobloch}, \citenamefont {Uecker},\ and\ \citenamefont
		{Yochelis}}]{knobloch2021origin}%
	\BibitemOpen
	\bibfield  {author} {\bibinfo {author} {\bibfnamefont {E.}~\bibnamefont
			{Knobloch}}, \bibinfo {author} {\bibfnamefont {H.}~\bibnamefont {Uecker}},\
		and\ \bibinfo {author} {\bibfnamefont {A.}~\bibnamefont {Yochelis}},\
	}\href@noop {} {\bibfield  {journal} {\bibinfo  {journal} {Physical Review
				E}\ }\textbf {\bibinfo {volume} {104}},\ \bibinfo {pages} {L062201} (\bibinfo
		{year} {2021})}\BibitemShut {NoStop}%
\end{thebibliography}
%

\end{document}